\PassOptionsToPackage{unicode}{hyperref}
\PassOptionsToPackage{hyphens}{url}
\documentclass[
]{article}
\usepackage{amsmath,amssymb}
\usepackage{lmodern}
\usepackage{iftex}
\ifPDFTeX
  \usepackage[T1]{fontenc}
  \usepackage[utf8]{inputenc}
  \usepackage{textcomp} 
\else 
  \usepackage{unicode-math}
  \defaultfontfeatures{Scale=MatchLowercase}
  \defaultfontfeatures[\rmfamily]{Ligatures=TeX,Scale=1}
\fi
\IfFileExists{upquote.sty}{\usepackage{upquote}}{}
\IfFileExists{microtype.sty}{
  \usepackage[]{microtype}
  \UseMicrotypeSet[protrusion]{basicmath} 
}{}
\makeatletter
\@ifundefined{KOMAClassName}{
  \IfFileExists{parskip.sty}{%
    \usepackage{parskip}
  }{
    \setlength{\parindent}{0pt}
    \setlength{\parskip}{6pt plus 2pt minus 1pt}}
}{
  \KOMAoptions{parskip=half}}
\makeatother
\usepackage{xcolor}
\IfFileExists{xurl.sty}{\usepackage{xurl}}{} 
\IfFileExists{bookmark.sty}{\usepackage{bookmark}}{\usepackage{hyperref}}
\hypersetup{
  hidelinks,
  pdfcreator={LaTeX via pandoc}}
\usepackage{longtable,booktabs,array}
\usepackage{calc} 
\usepackage{etoolbox}
\makeatletter
\patchcmd\longtable{\par}{\if@noskipsec\mbox{}\fi\par}{}{}
\makeatother
\IfFileExists{footnotehyper.sty}{\usepackage{footnotehyper}}{\usepackage{footnote}}
\makesavenoteenv{longtable}
\usepackage{graphicx}
\makeatletter
\def\maxwidth{\ifdim\Gin@nat@width>\linewidth\linewidth\else\Gin@nat@width\fi}
\def\maxheight{\ifdim\Gin@nat@height>\textheight\textheight\else\Gin@nat@height\fi}
\makeatother
\setkeys{Gin}{width=\maxwidth,height=\maxheight,keepaspectratio}
\makeatletter
\def\fps@figure{htbp}
\makeatother
\usepackage[normalem]{ulem}
\ifLuaTeX
  \usepackage{selnolig}  
\fi

\author{
    Sonish Sivarajkumar, MS\textsuperscript{1}\textsuperscript{,2}\thanks{Present address: School of Computing and Information, University of Pittsburgh, Pennsylvania, PA, USA. Work was done while at Molecular Robotics, Kerala, India.},
    Pratyush Tandale, MS\textsuperscript{3}, 
    Ankit Bhardwaj, BS\textsuperscript{4}, \\
    Kipp W. Johnson, MD,PhD\textsuperscript{5}, 
    Anoop Titus, MD\textsuperscript{6}, 
    Benjamin S. Glicksberg, PhD\textsuperscript{7},\\
    Shameer Khader, PhD, MPH\textsuperscript{8}\textsuperscript{\dag}, 
    Kamlesh K. Yadav, PhD\textsuperscript{9, 10}\textsuperscript{\dag}, \\
    Lakshminarayanan Subramanian, PhD\textsuperscript{4}\thanks{Corresponding authors: shameer.khader20@imperial.ac.uk, kamlesh.yadav@tamu.edu, lakshmi@cs.nyu.edu}
}
\date{\textsuperscript{\dag}Corresponding authors}

\date{\textsuperscript{1}Molecular Robotics, Kerala, India; \
      \textsuperscript{2}School of Computing and Information, University of Pittsburgh, Pennsylvania, PA, USA; \
      \textsuperscript{3}Health Informatics \& Data Science, Georgetown University, Washington DC, USA; \
      \textsuperscript{4}Department of Computer Science, Courant Institute of Mathematical Sciences, New York University, New York, NY, USA; \
      \textsuperscript{5}Institute for Next Generation Healthcare, Mount Sinai Health System, New York, NY, USA; \
      \textsuperscript{6}Department of Preventive Cardiology, DeBakey Heart \& Vascular Center, Houston Methodist, Houston, TX, USA; \
      \textsuperscript{7}Hasso Plattner Institute for Digital Health, Icahn School of Medicine at Mount Sinai, New York, NY, USA; \
      \textsuperscript{8}Faculty of Medicine, Imperial College London, London, UK; \
      \textsuperscript{9}School of Engineering Medicine,  Texas A\&M University, Houston, TX, USA; \      
      \textsuperscript{10}Department of Translational Medical Sciences, Center for Genomic and Precision Medicine, Texas A\&M University, Houston, TX, USA; \
}

\usepackage[left=3cm,right=3cm,top=3cm,bottom=3cm]{geometry}

\begin{document}

\title{Generation of a Compendium of Transcription Factor Cascades and Identification of Potential Therapeutic Targets using Graph Machine Learning }

\maketitle 

\textbf{Abstract}

Transcription factors (TFs) play a vital role in the regulation of gene expression thereby making them critical to many cellular processes. In this study, we used graph machine learning methods to create a compendium of TF cascades using data extracted from the STRING database. A TF cascade is a sequence of TFs that regulate each other, forming a directed path in the TF network. We constructed a knowledge graph of 81,488 unique TF cascades, with the longest cascade consisting of 62 TFs. Our results highlight the complex and intricate nature of TF interactions, where multiple TFs work together to regulate gene expression. We also identified 10 TFs with the highest regulatory influence based on centrality measurements, providing valuable information for researchers interested in studying specific TFs. Furthermore, our pathway enrichment analysis revealed significant enrichment of various pathways and functional categories, including those involved in cancer and other diseases, as well as those involved in development, differentiation, and cell signaling. The enriched pathways identified in this study may have potential as targets for therapeutic intervention in diseases associated with dysregulation of transcription factors. We have released the dataset, knowledge graph, and graphML methods for the TF cascades, and created a website to display the results, which can be accessed by researchers interested in using this dataset. Our study provides a valuable resource for  understanding the complex network of interactions between TFs and their regulatory roles in cellular processes. 

Website:
\href{https://tfcascades.streamlit.app/}{\uline{https://tfcascades.streamlit.app/}}

\hypertarget{introduction}{%
\subsection{Introduction:}\label{introduction}}

A Transcription Factor (TF) is a protein that regulates the
transcription of genetic information from DNA to RNA. Transcription
factors bind to specific DNA sequences, called response elements, and
help to control the expression of particular genes. They can either
activate or inhibit the transcription of a gene, depending on the
specific TF and the cell type in which it is functioning. TFs are
important for the proper regulation of gene expression and are involved
in many cellular processes, including development, cell growth and
division, and response to environmental signals\cite{Lambert2018TheFactors}.The key component of TFs
is that they possess a DNA-binding domain and directly bind to DNA,
which distinguishes them from other proteins, such as kinases,
methylases, co-activators, histone deacetylases, histones
acetyltransferases, and chromatin remodelers, , that lack this domain
and act indirectly on gene expression. TFs are involved in regulating
important pathways, such as immune responses\cite{Lee2020GlobalStudy}  and cell type
specification\cite{Lee2013TranscriptionalDisease}, and are used in laboratory experiments for cell
differentiation\cite{Fong2013SkeletalRe-programming}, de-differentiation, and trans-differentiation\cite{Takahashi2016APluripotency}.

TFs can regulate gene expression by themselves or in cascades by
activating the expression of a second TF, which then activates a third
TF, resulting in the amplification of the original signal\cite{Naika2013STIFDB2:Rice}\cite{Phillips2008TheExpression} (Figure 1).For this, the TF binds to specific DNA sequences and regulates another TF. This activated TF in turn goes on to regulate a third TF, creating cascades of gene expression. Thus, TF cascades refer to a series of TFs that work together to regulate gene expression. This cascade of events that leads to the regulation of a particular gene or set of genes results in the amplification of the initial signal and provides a regulatory relationship among TFs to maintain a high level of control over the expression of the target gene. 

\includegraphics[width=5.55446in,height=2.96297in]{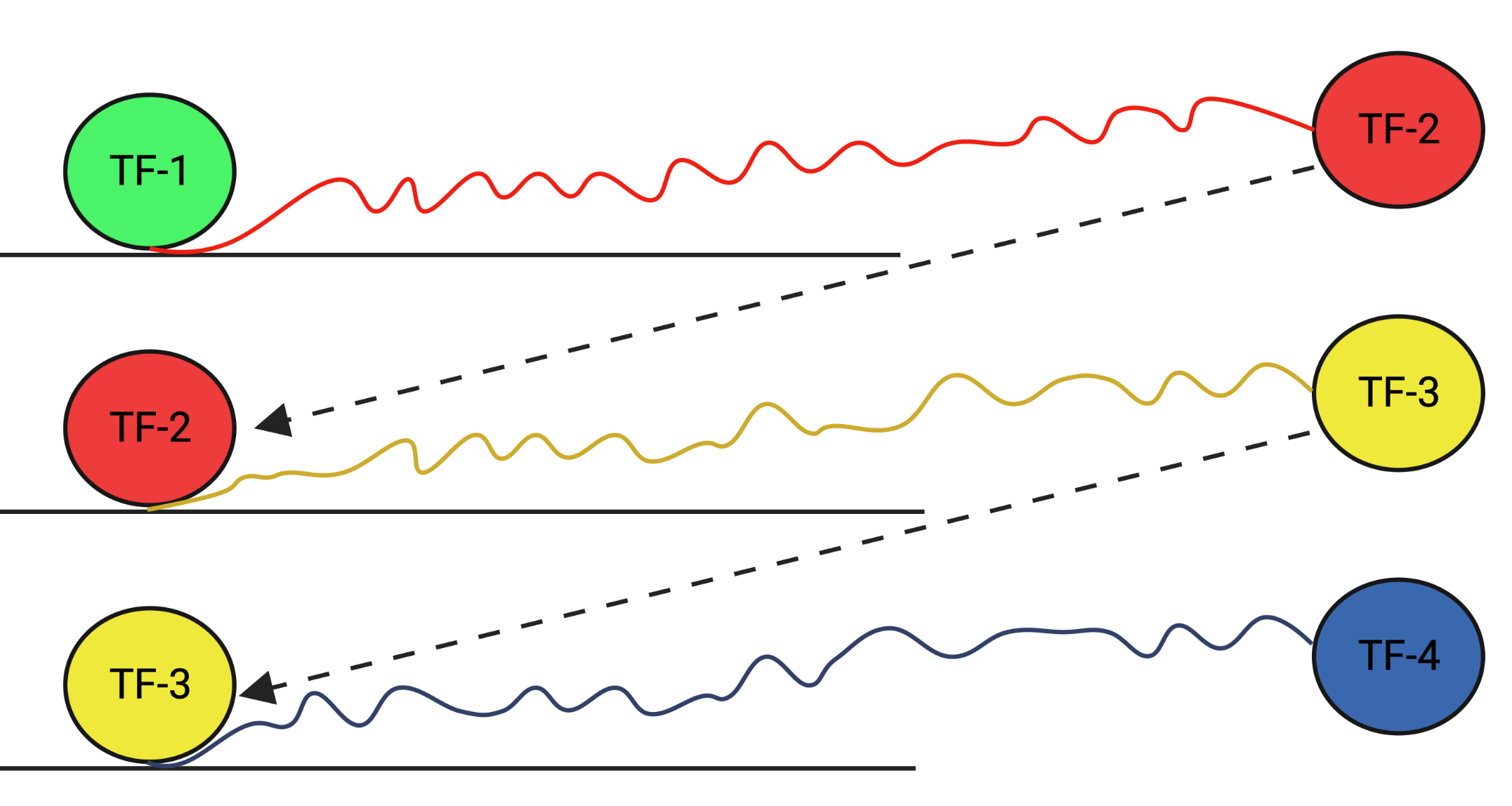}

\emph{\textbf{Figure 1 -} TF-1 binding to the DNA and activating TF-2
which induced the expression of TF-3 which in turn regulates TF-4
resulting in a cascade of TF production}

TF cascades can be disrupted by mutations in the genes that encode the TFs or by changes in the regulatory elements that control their expression. Dysregulation of TF cascades can lead to abnormal gene expression and can contribute to the development of diseases such as cancer. For example,  one of the most commonly known and most frequent mutated TF  in cancer is  TP53\cite{Alvarez2020TransientCascade}.

Thus, TF cascades are important for the proper regulation of gene
expression and are involved in many cellular processes, including
development, cell growth and division, and response to environmental
signals. They allow cells to respond to specific signals in a highly
coordinated and controlled manner. For example, when a cell receives a
signal from its environment, a TF cascade may be activated, leading to
the expression of specific genes that allow the cell to respond to the
signal.

TF cascades can be disrupted by mutations in the genes that encode the
TFs or by changes in the regulatory elements that control their
expression. Dysregulation of TF cascades can lead to abnormal gene
expression and can contribute to the development of diseases such as
cancer. Mutations in TFs can cause cancer as in the case of one of the
most commonly known and most frequent mutations of a TF TP53\cite{Surget2013UncoveringPerspective}. There
have been numerous clinical data records showing that TP53 is found to
be mutated in the majority of types of cancer as shown in figure 2.

\includegraphics[]{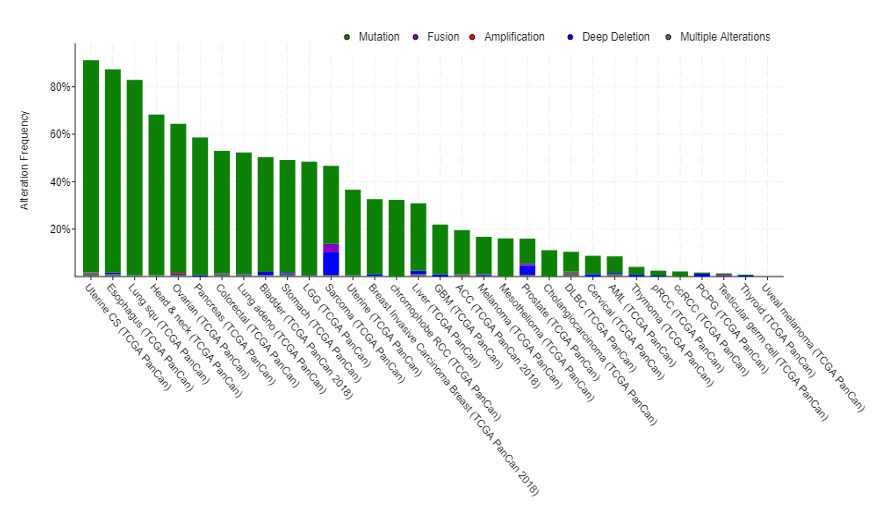}

\emph{\textbf{Figure 2} -- Alteration Frequency percentage vs types of
cancer bar graph plot for TP53 gene in human (developed using data from
cBioPortal\cite{Cerami2012TheData} )}

Thus, despite their potential role in cancer therapy, genome-scale TF cascades have remained largely unexplored and underutilized. In fact, due to the lack of information and appreciation for their role in cancer treatment till date, not many studies have identified TF cascades as therapeutic targets. This has created a significant knowledge gap and a missed therapeutic opportunity.  Therefore, there is an urgent need to compile a comprehensive and high-quality resource on genome-scale TF cascades and their associated pathways, which can facilitate the discovery of novel drug targets and biomarkers. Moreover, such a resource can enable the
application of advanced analytics and AI approaches, such as network
analysis, knowledge graphs, and graph machine learning, which can
capture the complex and dynamic interactions among TFs and genes, and
provide more accurate and robust predictions than conventional
methods\cite{Karamouzis2011TranscriptionParadigm}. A resource on genome-scale TF cascades can thus
revolutionize the field of cancer genomics and therapeutics.

Our study leveraged multiple techniques, including network analysis, concepts from knowledge graphs, and graph machine learning, to analyze TF cascades and their role in various diseases. Specifically, we began with TF-pathway enrichment analysis to identify relevant pathways and generate a list of candidate genes. Next, we constructed a comprehensive representation of the interactions between TFs and human biological pathways. To achieve this goal, we incorporated data from multiple sources and constructed a TF cascades graph to capture the interactions within TFs and between TFs and biological pathways. Through the application of network analysis and graph machine learning techniques, we were able to identify key regulatory relationships and predict novel interactions, enabling us to uncover potential therapeutic targets and biomarkers for further exploration. Overall, our study highlights the power of integrating multiple approaches to tackle complex biological problems and provides a valuable resource for researchers working in the field of drug discovery and precision therapeutics.

The use of graph-based methods has emerged as a promising approach for
predicting drug-target interactions and therapeutic interactions with
biomarkers\cite{Zhang2017Network-basedOncology}. Recent studies have demonstrated that graph database methods using knowledge graphs and graph machine learning can be used to represent biological entities and relationships, and enable entity-target interaction studies\cite{Olayan2018DDR:Approaches, Thafar2022Affinity2Vec:Learning}. Our proposed TF cascades graph and dataset provide a comprehensive and high-quality resource for researchers to explore the intricate interactions between TFs and genes, leading to the identification of new therapeutic targets and biomarkers. By integrating deep learning and graph-based approaches, we can overcome the limitations of previous methods and gain more profound insights into the underlying biology of diseases. Consequently, the impact of our TF cascades graph and dataset will be far-reaching, facilitating drug discovery and development, and ultimately improving patient outcomes.

\hypertarget{data-and-methods}{%
\subsection{\texorpdfstring{Data and Methods
}{Data and Methods }}\label{data-and-methods}}

\hypertarget{data}{%
\subsubsection{2.1 Data}\label{data}}

\hypertarget{string-database}{%
\subparagraph{STRING Database:}\label{string-database}}

STRING (Search Tool for the Retrieval of Interacting Genes/Proteins) is
a database of protein-protein interactions that includes both physical
interactions (such as those mediated by direct physical contact between
proteins) and functional associations (such as those mediated by shared
protein function or shared regulation of gene expression)\cite{Szklarczyk2016TheAccessible}. The
database includes interactions from a wide variety of sources, including
high-throughput experiments, computational predictions, and manual
curation from the scientific literature.

STRING is a useful resource for researchers who are interested in
understanding the relationships between proteins and how they function
in cells. It can be used to identify potential protein-protein
interactions, predict the functions of proteins, and understand how
proteins function in the context of larger networks. The database is
also useful for studying the relationships between proteins and
diseases, as disruptions in protein-protein interactions can contribute
to the development of diseases such as cancer.

In addition to protein-protein interactions, STRING also includes
information on small molecules and their interactions with proteins. It
also includes functional annotation for proteins, including information
on their pathways, functions, and structural features. The database is
constantly being updated with new information, and it is widely used by
researchers in the field of biology and biomedicine.

\hypertarget{tissues-database}{%
\subparagraph{TISSUES Database:}\label{tissues-database}}

TISSUES is a weekly updated web resource that integrates evidence on
tissue expression from manually curated literature, proteomics and
transcriptomics screens, and automatic text mining\cite{Palasca2018TISSUESExpression}. They map all
evidence to common protein identifiers and Brenda Tissue Ontology terms,
and further unify it by assigning confidence scores that facilitate
comparison of the different types and sources of evidence.

\hypertarget{cbioportal}{%
\subparagraph{cBioPortal:}\label{cbioportal}}

cBioPortal is a database which contains patient cancer data, including clinical and genomic data\cite{Cerami2012TheData}. It has complete details about patients in a complete and modular fashion, where you can choose to download specific parts of the patent data i.e. genes getting mutated data or patient race or patient age, etc.

\hypertarget{methods}{%
\subsubsection{2.2 Methods}\label{methods}}

\hypertarget{tf-interactions-extraction}{%
\paragraph{2.2.1 TF interactions
extraction}\label{tf-interactions-extraction}}

The STRING database was used to obtain information on all protein interactions. This database encompasses all known interactions between approximately 5090 species. For the analysis, only the Human action data from version 11 of the database was extracted. To focus solely on transcription factors (TFs) among all the proteins, a list of TFs was required. This list, which was published earlier\cite{Lambert2018TheFactors} and  currently hosted on the University of Toronto's server. Subsequently, the interaction data of TFs were gathered by selecting known TFs from the Human protein dataset.

The TFs list had ENSEMBLE IDs; hence, it was necessary to convert these
into protein-stable IDs for comparison with our protein interactions.
UniProt IDs are unique and unambiguous, making them useful for comparing
and integrating data across different sources and databases\cite{Consortium2019UniProt:Knowledge}. Biomart
is a very useful tool that allows converting gene identifiers to
protein-stable IDs and it's a powerful tool for simplifying and
automating the process. We used the Biomart tool to convert these IDs to
protein-stable IDs by cross-referencing the IDs. We used the Biomart tool to convert these IDs to protein-stable IDs by cross-referencing the IDs. It was done by selecting the Human Protein database on the biomart website, selecting the necessary attributes such as gene name, and using a lookup function in the database to match the gene names in our list of TF interactions with the corresponding UniProt IDs in the Biomart dataset. The selection of the Human Protein database ensured that the extracted TF interactions only correspond to Human proteins and other interactions are filtered out.

Tissue data was extracted to know the location of each TF. The data was acquired from the TISSUES database. In this study, we have limited our study to cancer studies from cBioPortal. The data was fetched using a script from the \textit{cgdsr} R package\cite{Jacobsen2015Cgdsr:CGDS}. This package has functions, which are used for querying the Cancer Genomics Data Server (CGDS,
\url{http://www.cbioportal.org/datasets}), which is hosted by the
computational Biology Center at MSKCC.

\hypertarget{building-the-tf-cascades-dataset}{%
\paragraph{\texorpdfstring{\emph{Building the TF Cascades
dataset}}{Building the TF Cascades dataset}}\label{building-the-tf-cascades-dataset}}

The process of preprocessing the TF interactions data was crucial to
ensure that the data is accurate and reliable when building TF cascades.
We removed duplicate TF interactions to eliminate any redundant
information that existed in the dataset. Additionally, we formatted the
data in a way that could be easily used for building the TF- cascades,
i.e. we organized the data so that it could be easily read and analyzed
by the algorithms that were used to build the cascades.

Once the interaction data was preprocessed, we obtained all the mutual
connections to build a cascade of the TFs. Similarly, all possible
combinations of TF-interactions were connected to form a TF cascade
dataset. Here, connected components are the groups of TFs that are
connected to one another through interactions, and they represent the TF
cascades. The obtained cascades can be filtered out based on the size,
or specific requirements to have a more specific and informative result.
We also reindexed the dataset by adding the cascade level to each of the
cascades. For instance, TF 1 -\/-\textgreater{} TF 2 -\/-\textgreater{}
TF 3 were assigned a length of 3, indicating 3 TFs are present in the
chain, but L2 as the cascade level as it has two levels of cascades.

\hypertarget{analytics}{%
\paragraph{\texorpdfstring{2.2.2 Analytics
}{2.2.2 Analytics }}\label{analytics}} 

\paragraph{
    \includegraphics[]{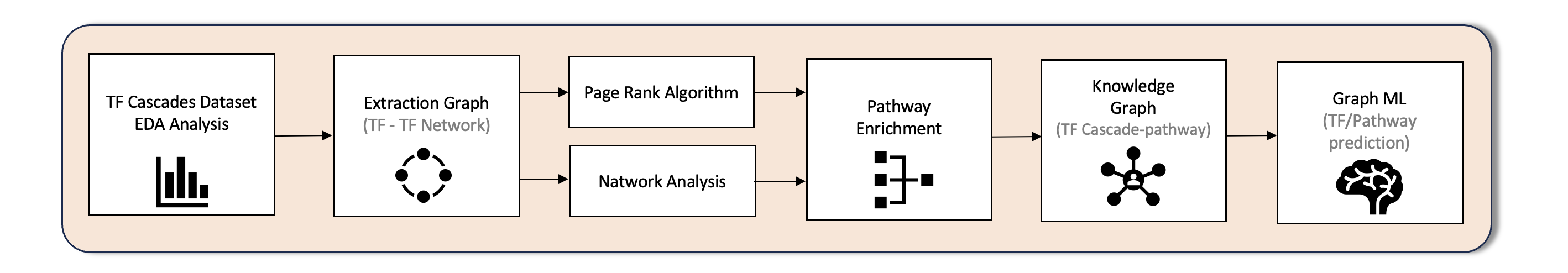}
    }

\emph{\textbf{Figure 3} -- Overall Analytics and ML workflow}

\hypertarget{cascade-summary}{%
\subparagraph{\texorpdfstring{Cascade summary
}{Cascade summary }}\label{cascade-summary}}

TF cascades refer to linear or branching pathways of gene regulation,
where the expression of one TF can influence the expression of
downstream TFs. We performed Exploratory Data Analysis (EDA) to assess
and summarize the database's primary characteristics, as it is difficult
to examine each element of a dataset and select the most relevant ones.
Our analysis was performed using the python library called
\textit{pandas-profiling} and employed different visualization methods to
discover the underlying patterns in the data, enabling us to comprehend
what the data is communicating. Several graphical and non-graphical
analysis techniques were employed to analyze the TF Cascades data. We
restricted our investigation to univariate analysis because we are
primarily interested in categorical variables - TFs.

\hypertarget{page-rank}{%
\subparagraph{\texorpdfstring{Page Rank }{Page Rank }}\label{page-rank}}

\textit{PageRank} is an algorithm developed by Google to rank web pages in their
search engine results based on the number and quality of links pointing
to them\cite{Page1998TheReport}. The algorithm is based on the idea that a web page is more
important if it is linked to by other important pages, and less
important if it is linked to by less important pages. PageRank has been
widely used and studied in the field of information retrieval and has
inspired the development of similar algorithms in other fields.

We applied the PageRank algorithm in the TF-Cascades dataset to rank the
TFs based on their connectivity and centrality within the cascade. TF
cascades can be represented as directed graphs, with TFs represented as
nodes and regulatory interactions represented as edges. 

To apply the PageRank algorithm to a gene cascade, the following steps
were followed:

\begin{enumerate}
\def\labelenumi{\arabic{enumi}.}
\item
  Construct a directed graph representation of the gene cascade, with
  TFs as nodes and regulatory interactions as edges.
\item
  Initialize the PageRank scores of all genes to 1.
\item
  Iteratively update the PageRank scores of all genes based on the
  PageRank scores of the genes that regulate them, using the following
  formula: PR(A) = (1 - d) + d * SUM(PR(T) / C(T)) where PR(A) is the
  PageRank score of gene A, PR(T) is the PageRank score of the genes
  that regulate gene A (T), C(T) is the number of genes regulated by T,
  and d is a damping factor that accounts for the possibility that the
  walk through the graph may not always follow the edges (usually set to
  0.85).
\item
  Repeat step 3 until convergence, i.e. until the PageRank scores of the
  TFs do not change significantly between iterations.
\end{enumerate}

The resulting PageRank scores were used to rank the genes in the gene
cascade based on their importance and centrality within the cascade. The
PageRank algorithm provided valuable insights into the functional roles
and relationships of the TFs in TF-cascade.

\hypertarget{extraction-graph}{%
\subparagraph{Extraction Graph}\label{extraction-graph}}

TF cascades can be represented as directed graphs, where the TFs are
represented as nodes and the regulatory interactions between them are
represented as edges. The direction of the edges reflects the direction
of the regulatory influence, with the upstream TFs regulating the
downstream TFs. This will facilitate a better understanding of the TFs
with the highest scores (represented by the graph's nodes). This will
allow us to focus on the most important TFs in terms of medication
development and personalized medicine.

For example, consider a simple TF cascade with three TFs, A, B, and C,
where A regulates B, and B regulates C. This cascade can be represented
as a directed graph with three nodes (A, B, C) and two edges (A
-\textgreater{} B, B -\textgreater{} C). The direction of the edges
reflects the regulatory relationships between the TFs, with A regulating
B and B regulating C. The graph representation of a TF cascade captures
the hierarchical structure and relationships between the TFs and can be
used to analyze and understand the regulatory mechanisms of the cascade.

The graph representation of a TF cascade can be constructed from
experimental or computational data on the regulatory interactions
between the TFs. There are various databases and resources available
that provide information on transcriptional regulatory interactions,
such as TRED and Transfac. In addition to the regulatory TF
interactions, other information, such as the associated biological
pathways, and timing of the regulatory pathways, can be incorporated
into the graph representation to provide a more comprehensive view of
the TF cascade, which is done in the later part of the study. The graph
representation of a TF cascade was analyzed using various network
analysis algorithms and tools to identify patterns and trends in the
regulatory network and to gain insights into the underlying regulatory
mechanisms.

\hypertarget{network-analysis-on-the-extraction-graph}{%
\subparagraph{\texorpdfstring{\emph{Network Analysis on the Extraction
Graph}}{Network Analysis on the Extraction Graph}}\label{network-analysis-on-the-extraction-graph}}

Network analysis is a powerful approach for exploring and understanding
the relationships between genes and pathways in the context of complex
biological systems\cite{Zhang2017Network-basedOncology, Shameer2018ADisease, Glicksberg2016ComparativeNetworks}. In the context of TF cascades, network analysis
can be performed on the directed graph representation of the TF cascade
to gain insights into the functional connections and interdependencies
between the TFs and the pathways of transcriptional regulation.

Centrality measures are a class of network analysis metrics that are
used to identify the most central or influential nodes in a network.
There are several types of centrality measures, including degree
centrality, betweenness centrality, and eigenvector centrality.

Degree centrality measures the number of connections a node has in the
network. In the context of a TF cascade, degree centrality can be used
to identify the TFs that have the most regulatory interactions with
other TFs. These TFs may be central to the functioning of the cascade
and may be potential therapeutic targets.

Betweenness centrality measures the number of shortest paths between
pairs of nodes that pass through a given node. In the context of a TF
cascade, betweenness centrality can be used to identify the TFs that are
most central or influential in the cascade, as they are likely to be
involved in multiple regulatory pathways and may have a broad impact on
the cascade.

Eigenvector centrality measures the influence of a node based on the
influence of its neighbors. In the context of a TF cascade, eigenvector
centrality can be used to identify the TFs that are most central or
influential in the cascade, based on the collective influence of their
regulatory targets. TFs with high eigenvector centrality are likely to
have a broad impact on the cascade and may be potential therapeutic
targets.

\hypertarget{pathway-enrichment}{%
\subparagraph{\texorpdfstring{Pathway enrichment
}{Pathway enrichment }}\label{pathway-enrichment}}

We conducted gene set enrichment analysis for the purpose of functional
investigation of TF cascades and identification of disease phenotypes.
Pathway enrichment analysis is a common approach used in bioinformatics
to identify pathways or biological processes that are significantly
enriched in a given gene set. This is often done to understand the
underlying biological mechanisms or functions associated with a
particular set of genes, such as those differentially expressed in a
disease state or in response to a particular treatment.

We used the Enrichr tool\cite{Kuleshov2016Enrichr:Update} to perform pathway enrichment analysis on a
set of 500 human TFs. Enrichr is a widely used online tool, which allows
users to input a list of genes and returns a list of significantly
enriched pathways and functional categories based on the gene set.
Enrichr uses a large number of databases and resources, including KEGG,
Reactome, Gene Ontology, and others, to provide a comprehensive coverage
of pathways and functional annotations.

Each TF cascade was compared to a reference set of genes (all genes in
the human genome), and the p-value was used to identify pathways that
are overrepresented in the TF set. The p-value is a statistical measure
that indicates the likelihood of obtaining a result as extreme or more
extreme than the one observed, given that the null hypothesis is true.
The p-value is commonly used to assess the statistical significance of a
result and to determine whether it is due to chance or reflects a true
effect. In the context of pathway enrichment analysis, the p-value is
used to determine the likelihood that the observed enrichment of a
particular pathway in the input gene set is due to chance, rather than
being a true reflection of the underlying biology.

The adjusted p-value is a modified version of the p-value that takes
into account the multiple comparisons problem, which occurs when
multiple tests are performed simultaneously and the probability of
obtaining a significant result by chance increases. The adjusted p-value
is calculated using methods such as the Bonferroni correction or the
Benjamini-Hochberg method\cite{Thissen2002QuickComparisons}, which control for the false discovery
rate (FDR) - the expected proportion of false positives among the
significant results. We used the adjusted p-value to prioritize the
pathways and functional categories that are most likely to be true
positives, rather than false positives due to multiple testing.The
Benjamini-Hochberg method is used to calculate the adjusted p-value,
which controls the FDR at the specified level (e.g. 0.05).

The results of pathway enrichment analysis could provide valuable
insights into the biological functions and mechanisms associated with
the extracted TF set\cite{Li2017ComprehensiveDatasets}.

\hypertarget{knowledge-graph-modeling}{%
\subparagraph{2.2.3 Knowledge graph
modeling}\label{knowledge-graph-modeling}}

A knowledge graph is a structured representation of knowledge that
captures the relationships between entities and their attributes. We
constructed a knowledge graph on the results of pathway enrichment
analysis to capture the relationships between the enriched pathways and
the TFs in the cascade. Figure 4 shows the steps in the knowledge graph
modeling.

\includegraphics[]{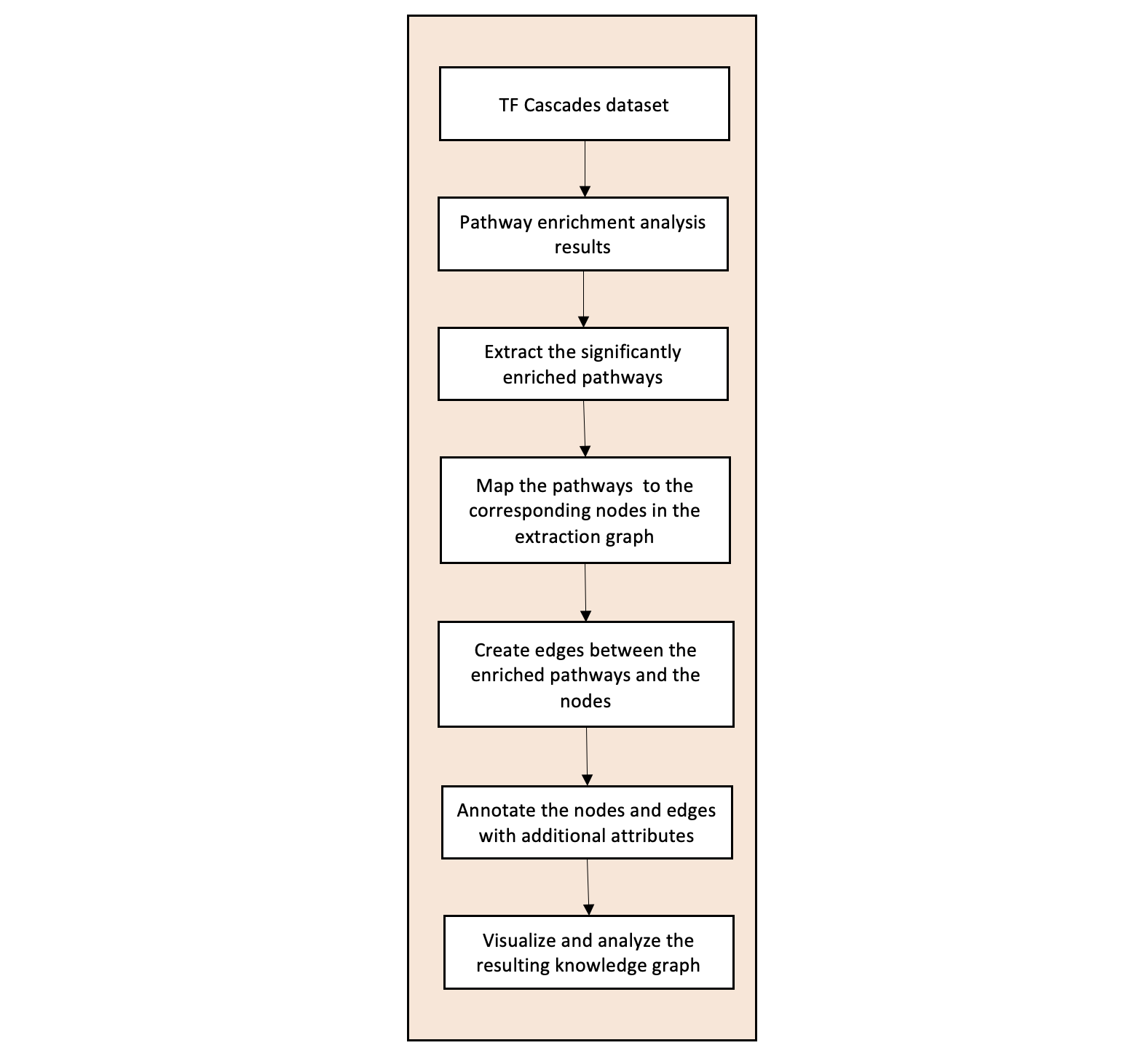}

\emph{\textbf{Figure 4} -- Knowledge graph modeling workflow}

From the pathway enriched TF cascades dataset, we identified the
significantly enriched pathways in the TF cascade, based on the adjusted
p-value threshold of 0.05. Next, we filtered the genes in the enriched
pathways and mapped them to the corresponding nodes in the directed
graph representation of the TF cascades. We modeled the edges between
the enriched pathways and the corresponding nodes in the directed graph
to capture the relationships between the pathways and the TFs. The edges
were created by linking the enriched pathways, represented as nodes, to
the corresponding TFs, represented as nodes in the directed graph
(extraction graph). The nodes and edges were then annotated with
additional attributes, such as gene names, pathway names, and regulatory
relationships\cite{Yacoumatos2021TrialGraph:Trials}. The nodes were annotated with the gene names and the
pathways were annotated with their names and descriptions. We also
embedded the cascade levels in each cascade as the attribute of the TF
nodes. Finally, we created a NetworkX visualization\cite{Hagberg2008ExploringNetworkX} of the resulting
knowledge graph to gain insights into the relationships between the
enriched pathways and the TFs in the cascade.

The knowledge graph on the TF cascades could provide a structured and
interactive representation of the relationships between the enriched
pathways and the TFs in the cascade. It was then used to explore and
analyze the relationships between the pathways and the TFs and to
understand the functional roles and relationships of the TFs within the
cascade. The knowledge graph can also be used to identify potential
therapeutic targets and mechanisms of action and to inform further
experimental studies.

\includegraphics[]{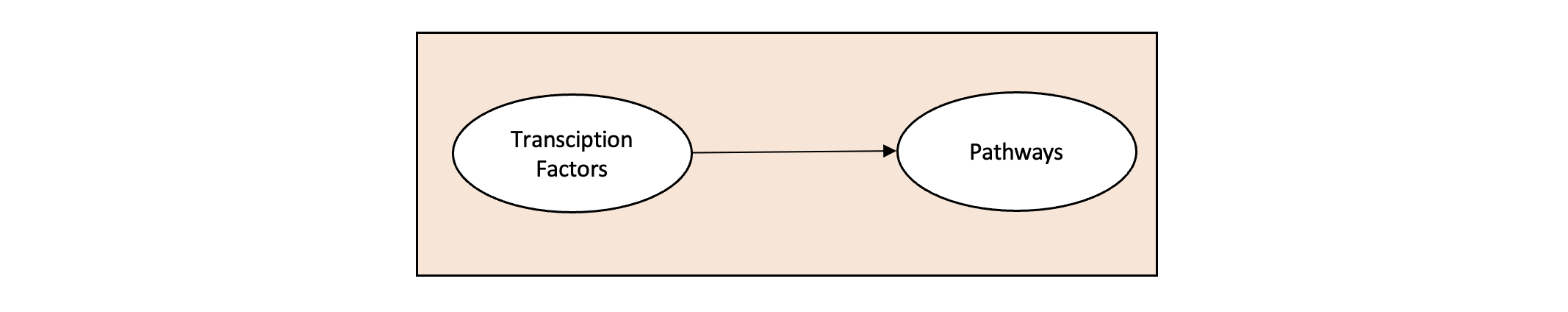}

\emph{\textbf{Figure 5}: Graph Schema}

The purpose of creating a graph from the newly constructed TF cascades
and enriched pathway dataset is to establish a more comprehensive and
generalizable representation of TF cascades. This approach enables us to
construct a foundational model that can be readily applied to various
downstream tasks, such as Next TF prediction, TF classification, and
Pathway prediction, among others. By doing so, we ensure that
researchers in this field have access to a complete resource that
includes the dataset as well as the foundational model. Consequently,
this approach can serve as a valuable tool to promote and advance
further research in this area. Moreover, graph approaches are preferred
over Recurrent Neural Network(RNN) and Hidden Markov Model(HMM): approaches when we are interested in the relationships
between entities rather than sequences of states of a single entity. The
TF cascades problem fits the graph paradigm better because different TFs
are different entities that have regulatory relationships between them,
and not a single entity showing different sequences of states.
Therefore, our graph-based model can capture the complex and dynamic
interactions among TFs more effectively than other methods.

\hypertarget{graph-machine-learninggraph-ml-tasks-that-enables-biological-discovery}{%
\subparagraph{2.2.4 Graph Machine Learning(Graph ML) tasks that enables
biological
discovery}\label{graph-machine-learninggraph-ml-tasks-that-enables-biological-discovery}}

Graph machine learning has emerged as a promising approach for analyzing
heterogeneous biological data by capturing complex relationships between
entities\cite{Gaudelet2021UtilizingDevelopment}. Common graph machine learning tasks include link
prediction, which can be used to predict protein-protein interactions or
drug-target interactions, and node classification, which can predict the
function of a protein or the type of a cell. Graph embeddings can also
be learned to provide low-dimensional vector representations for each
node in the graph, which can be used for downstream tasks\cite{Makarov2021SurveyGraphs}.

\includegraphics[]{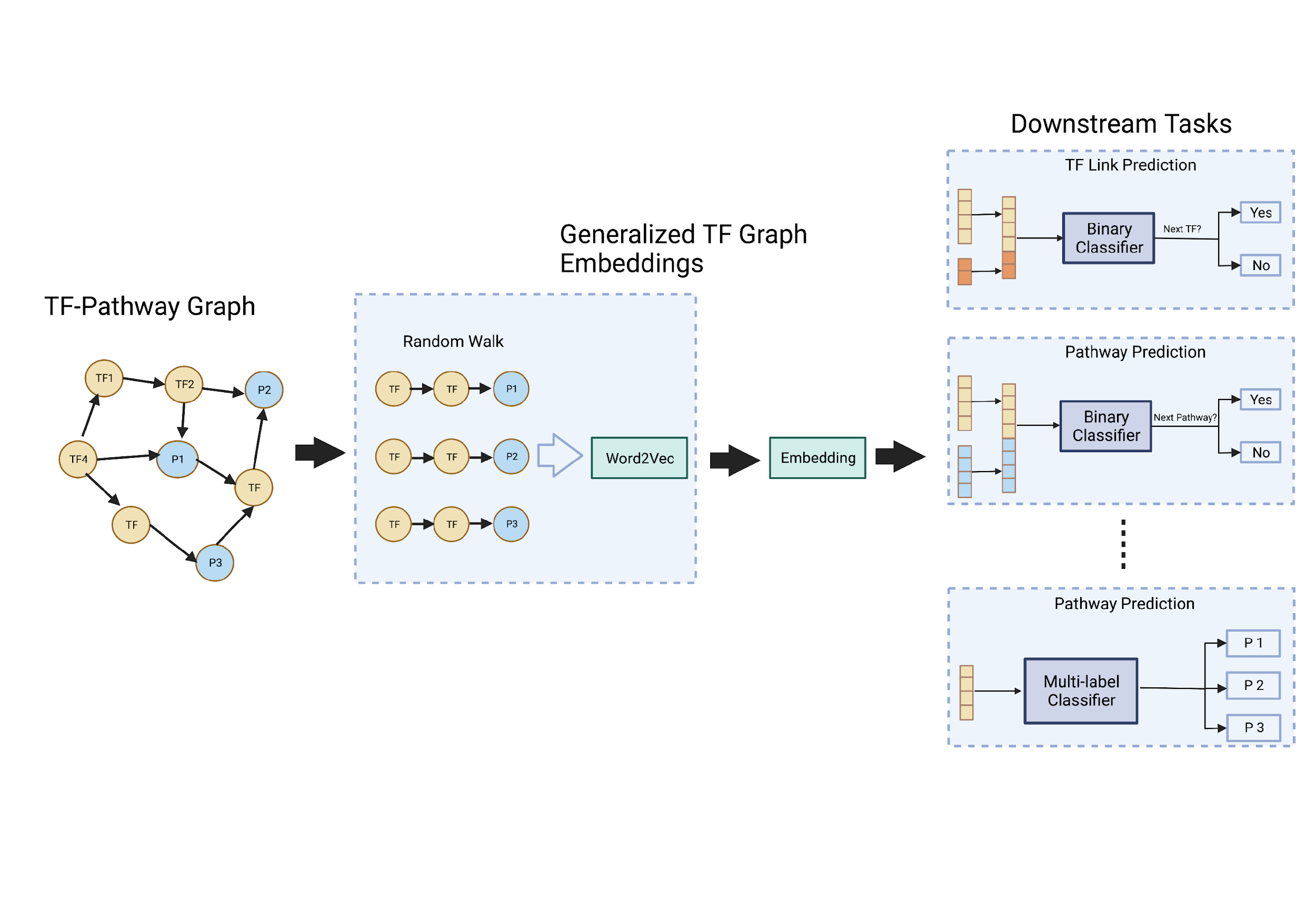}

\begin{quote}
\emph{\textbf{Figure 6:} Node2vec based graph embedding workflow
diagram}
\end{quote}

\textbf{TF prediction:}

Link prediction is a technique that can be applied to predict missing or
future links in a network, such as relationships between genes or
pathways in a biological network. We employed link prediction methods to
predict the next TF in a given cascade, based on the relationships
between the existing TFs in the cascade. This approach is based on the
idea that nodes with similar attributes are more likely to be linked in
the future. The cascade levels of TFs in the cascade, which were used as
attributes for the nodes, were considered as the expression level of a
TF and a similarity measure (Cosine similarity), was applied.

For performing this similarity-based TF link prediction in a given
cascade, we employed a network embedding technique called
Node2vec{[}17{]} (Figure 6). The Node2vec algorithm was used to learn
low-dimensional representations of the nodes in the enriched TF cascades
KG, which could capture the network's topology and could preserve the
structural relationships between nodes. The cosine similarity measure
was then applied to the learned TFCascade-pathway embeddings to predict
the next TF in a given cascade. It is important to note that Node2vec
can handle large and sparse networks and can also capture higher-order
relationships, which makes it suitable to be used as a TF link
prediction method.

To obtain a link embedding from the node embeddings, we applied
different operators such as Hadamard, L1, L2, and average to combine
them. These operators are applied to combine the embeddings of the two
nodes to form the resulting link embedding. The hadamard operator, also
known as the element-wise multiplication operator, takes the
element-wise product of the embeddings of two nodes, resulting in a link
embedding with the same dimensionality as the input node embeddings.
Mathematically, for two embeddings u and v, the Hadamard operator is
defined as $u \odot v$ .

The L1 operator, also known as the Manhattan distance operator, takes
the absolute difference between the embeddings of two nodes, resulting
in a link embedding with the same dimensionality as the input node
embeddings. Mathematically, for two embeddings u and v, the L1 operator
is defined as \textbar u - v\textbar. The L2 operator, also known as the
Euclidean distance operator, takes the square root of the sum of the
squared differences between the embeddings of two nodes, resulting in a
link embedding with the same dimensionality as the input node
embeddings. Mathematically, for two embeddings u and v, the L2 operator
is defined as $\sqrt{(u - v)^2}$.

The average operator takes the average of the embeddings of two nodes,
resulting in a link embedding with the same dimensionality as the input
node embeddings. Mathematically, for two embeddings u and v, the average
operator is defined as (u + v)/2. These operators have different
implications for link prediction performance, depending on the
characteristics of the graph and the specific task at hand. The next
step was to evaluate and rank the predicted links based on the scores
generated by the link prediction algorithm.

\includegraphics[]{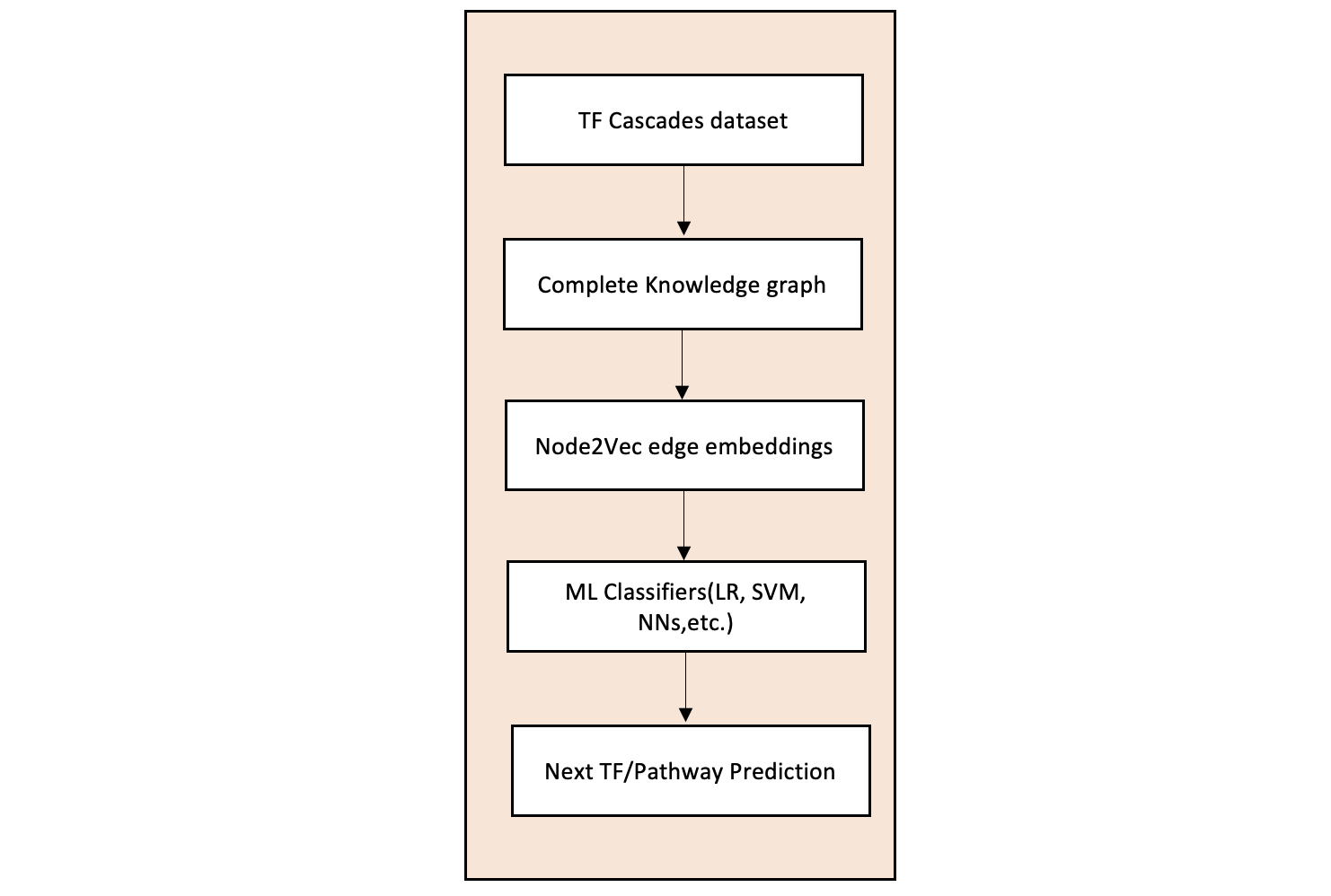}

\emph{\textbf{Figure 7:} Overall workflow of using TF Cascades for next
TF or pathway prediction}

Pathway prediction is another experiment which we performed using the
pathway-enriched TF-Cascades knowledge graph. The process can be similar
to the steps used for predicting the next TF in a cascade, but instead
of predicting links between TFs, the focus would be on predicting links
between pathways and TFs.

TF-Cascades link prediction is a powerful approach but it can not
replace experimental validation and functional assays as it would be
crucial to validate the TF predictions experimentally, to ensure their
accuracy, especially for therapeutic applications. It's important to
mention that link prediction is a challenging task, and the performance
of link prediction methods can be affected by the sparsity and the size
of the network. It's also crucial to validate the predictions
experimentally. The TF or pathway predictions can be tested by
perturbing the genes in the network using gene editing techniques or
other methods, and observing the changes in the network's topology.

In addition to next TF or pathway prediction, the following approaches
can also be used to identify potential therapeutic targets using TF
cascades dataset and knowledge graph:

\begin{itemize}
\item
  Identifying Key TFs: The knowledge graph can be analyzed to identify
  the TFs that are central or influential in the cascade, based on their
  connectivity and centrality within the network. These TFs may be
  central to the functioning of the cascade and may be potential
  therapeutic targets. For example, by using centrality measures like
  degree centrality, betweenness centrality and eigenvector centrality
  on the knowledge graph one can identify the TFs that have the most
  regulatory interactions with other TFs, or the TFs that are most
  central or influential in the cascade, as they are likely to be
  involved in multiple regulatory pathways and may have a broad impact
  on the cascade.
\item
  Identifying Enriched Pathways: The knowledge graph can be analyzed to
  identify the enriched pathways that are most significant or relevant
  to the cascade, based on their p-value or other statistical measures.
  These pathways may be involved in the cascade and may be potential
  therapeutic targets.
\item
  Network Analysis: One can use various network analysis techniques like
  clustering, modularity, centrality etc. to identify the modules and
  subnetworks of the knowledge graph, which may represent the functional
  pathways and mechanisms of action within the cascade.
\item
  Identifying Regulatory Relationships: The knowledge graph can be
  analyzed to identify the regulatory relationships between the TFs and
  the pathways, and the mechanisms of these relationships. For example,
  by looking at the edges and annotation on the edges in the knowledge
  graph, one can identify the type of regulatory relationship (e.g.
  activation, repression) and any additional information about the
  regulatory mechanisms (e.g. binding sites, motifs)
\item
  Validation: The potential targets and mechanisms identified using the
  knowledge graph can be validated using experimental data. This can be
  done by performing experiments such as gene knockdown or
  overexpression, or by analyzing gene expression data from relevant
  biological contexts. Validation can provide valuable insights into the
  biological functions and mechanisms of action of the potential
  targets.
\end{itemize}

\hypertarget{results}{%
\subsection{3. RESULTS}\label{results}}

\hypertarget{tf-cascades-dataset}{%
\subsubsection{3.2.1 TF Cascades dataset:}\label{tf-cascades-dataset}}

In this study, we aimed to create a compendium of TF cascades from the
dataset of TF interactions extracted from the STRING database. The raw
human protein interaction data obtained from the STRING database
contained a total of 11 million interactions. To focus specifically on
transcription factors, the dataset was pruned by removing all non-TF
proteins, resulting in a reduced dataset of 300,000 interactions.

The whole protein ID TFs cascade was converted to gene IDs TFs cascades by creating a dictionary using BioMart where one column contained all the Protein IDs and the second column contained their respective Gene name as shown in Table 1.  The TFs Cascades were converted to gene cascades using python script.

\emph{\textbf{Table 1}: Dictionary containing Protein ID and Gene name
pairs}

\begin{longtable}[]{@{}
  >{\raggedright\arraybackslash}p{(\columnwidth - 2\tabcolsep) * \real{0.5000}}
  >{\raggedright\arraybackslash}p{(\columnwidth - 2\tabcolsep) * \real{0.5000}}@{}}
\toprule
\endhead
Protein stable ID & Gene name \\
\bottomrule
ENSP00000177694 & TBX21 \\
ENSP00000201031 & TFAP2C \\
ENSP00000204517 & TFAP4 \\
ENSP00000216037 & XBP1 \\
ENSP00000217026 & MYBL2 \\
ENSP00000217086 & SALL4 \\
ENSP00000221452 & RELB \\
ENSP00000222122 & DBP \\
ENSP00000222598 & DLX5 \\
ENSP00000222726 & HOXA5 \\
\bottomrule
\end{longtable}

Tissue data was now taken and it was filtered to keep only proteins of interest.
Cancer Patient Data was taken from cBioPortal and genes which showed mutation(s) were cross checked with the genes which were present in our cascades. From the TF cascades built we identified 81,488 unique TF cascade chains, where the longest chain had a length of 62 TFs (cascade level L61, Figure 8) having only 426 unique genes out of 1636 total present in humans. Table 3. The cascades were plotted according to different lengths (Figure 8) where cascades having length 43 had the highest frequency of 2815. We also found more than 2500 occurrences for TF cascade levels from L39 to L43.
 
These results indicate a complex network of interactions between TFs, and suggest that multiple TFs work together to regulate gene expression. The high number of unique TF cascades (81,488) highlights the intricate nature of these interactions. Furthermore, the length of the longest cascade (L61) indicates the scale of these regulatory networks.

\includegraphics[width=6.10887in,height=3.15625in]{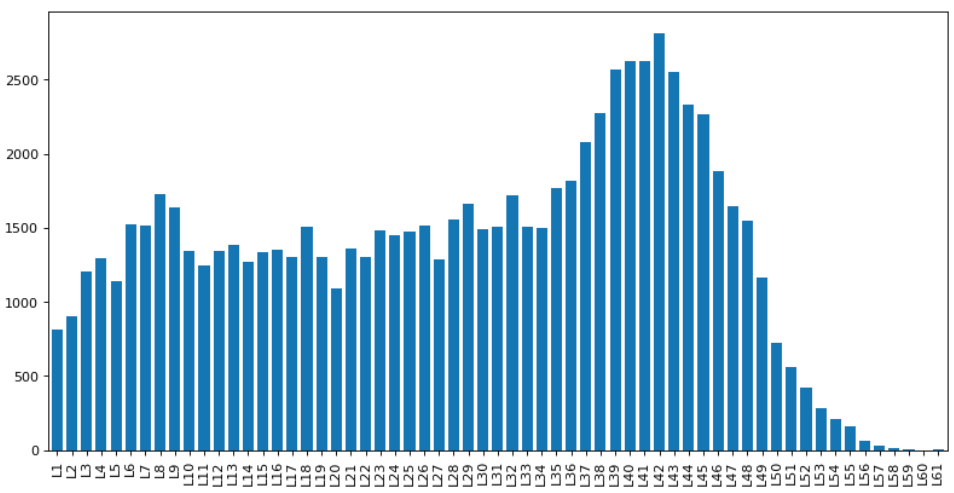}

\emph{\textbf{Figure 8}: Frequency distribution of TF cascades in each
cascade level}

\hypertarget{a-summary-of-tf-cascades-dataset}{%
\subsubsection{3.2.2 a) Summary of TF Cascades
dataset:}\label{a-summary-of-tf-cascades-dataset}}

We created a website to show the EDA results to researchers interested
in using the TF Cascades
dataset(\href{https://sonishsivarajkumar.github.io/TFCascades/}{\emph{\uline{https://sonishsivarajkumar.github.io/TFCascades/}}}).

\includegraphics[width=6.08654in,height=3.10542in]{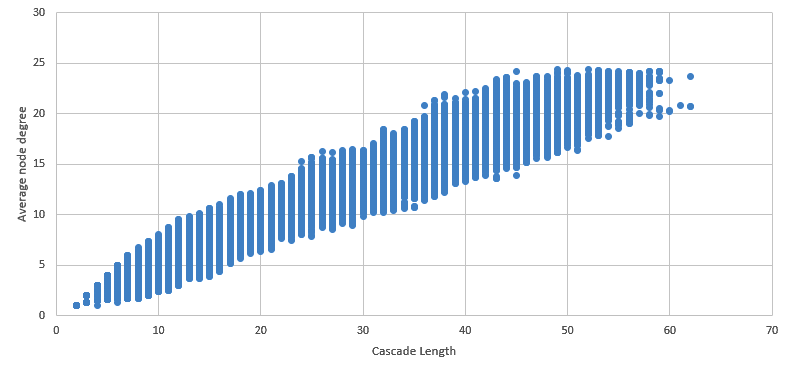}

\emph{\textbf{Figure 9} - Dot plot displaying average node degree vs
length of the cascade.}

The Exploratory Data Analysis (EDA) of the TF cascades dataset provided
valuable insights into its primary characteristics. The dataset consists
of 64 variables( cascade id, cascade level and 62 TFs) and 81,488
observations.

The index variable is a real number that assigns a cascade ID, ranging
from 1 to 81,488. It is essential for indexing the cascades and
assigning cascade levels. The "Level" variable, which represents the
cascade levels, is a categorical variable with 61 distinct values,
indicating that the cascades have different levels of complexity. The
most frequent cascade levels are Level 42, Level 41, Level 40, Level 39,
and Level 43. These levels are present in 2,815, 2,628, 2,625, 2,570,
and 2,548 cascades, respectively. The remaining cascade levels have a
total count of 68,302 cascades.

The "TF 1" variable represents the first TF in each cascade and also
exhibits high cardinality. It has 311 distinct values, with STAT3 being
the most frequent TF present in 3,419 cascades. Other frequently
occurring TFs include POU5F1 (1,546 cascades), TP53 (1,371 cascades),
SOX2 (1,166 cascades), and FOXO3 (1,157 cascades). The remaining TFs
have a total count of 72,829 cascades.Similarly, the "TF 2" variable
represents the second TF in each cascade and has 291 distinct values.
The most frequent TF in this variable is MYC, present in 4,231 cascades.
TP53 (2,821 cascades), NANOG (2,687 cascades), AR (2,091 cascades), and
FOXP3 (2,050 cascades) are also highly occurring TFs. The remaining TFs
have a total count of 67,608 cascades.

The 'TF 3' variable, representing the third TF in each cascade, has 233
distinct values. The most frequent TF in this variable is NANOG, present
in 12,659 cascades. SNAI1 (8,178 cascades), TP53 (4,448 cascades),
POU5F1 (4,385 cascades), and PPARG (2,670 cascades) are also prominent
TFs. There are 813 missing values in this variable. The remaining TF
variables (TF 4 to TF 62) exhibit a similar pattern, with varying
numbers of distinct TFs and missing values. These variables have high
cardinality and missing values ranging from 2.1\% to 39.1\%.

The EDA provided a comprehensive summary of the dataset, highlighting
the characteristics of the TF cascades. The dataset contains a diverse
set of TFs, each involved in various cascade levels. However, missing
values and high cardinality pose challenges in the analysis and
interpretation of the data. Further investigations and statistical
techniques were applied to uncover additional insights from this TF
cascades dataset.

\hypertarget{b-pagerank-results}{%
\subsubsection{3.2.2 b) PageRank Results}\label{b-pagerank-results}}

After running 10 iterations of PageRank with cascade connection serving
as the initial weights of all nodes set to 0.25, the analysis revealed
that the NANOG gene obtained the highest ranking with a score of
0.016963, while the tp53 gene was ranked second with a score of
0.011583.

\emph{\textbf{Table 2:} Result of PageRank}

\begin{longtable}[]{@{}
  >{\raggedright\arraybackslash}p{(\columnwidth - 4\tabcolsep) * \real{0.3687}}
  >{\raggedright\arraybackslash}p{(\columnwidth - 4\tabcolsep) * \real{0.3594}}
  >{\raggedright\arraybackslash}p{(\columnwidth - 4\tabcolsep) * \real{0.2719}}@{}}
\toprule
\endhead
TF & PageRank & Rank \\
\bottomrule
NANOG & 0.0169 & 1 \\
TP53 & 0.0115 & 2 \\
MYC & 0.0113 & 3 \\
AR & 0.0100 & 4 \\
SP7 & 0.0091 & 5 \\
SNA12 & 0.0085 & 6 \\
SNAI1 & 0.0083 & 7 \\
PPARG & 0.0082 & 8 \\
NEUROG3 & 0.0081 & 9 \\
RUNX2 & 0.0079 & 10 \\
\bottomrule
\end{longtable}

\hypertarget{c-network-analysis-of-the-extraction-graph}{%
\subsubsection{3.2.2 c) Network Analysis of the Extraction
Graph}\label{c-network-analysis-of-the-extraction-graph}}

The TF cascades extraction graph was built by combining all the cascades
into a single graph. The network comprised 426 nodes representing 426
unique TFs, and 866 edges which correspond to 866 interactions between
all the TFs. From the detailed analysis of the network, we identified 10
TFs which had the highest regulatory influence through three centrality
measurements -- Betweenness centrality, closeness centrality, and
eigenvector centrality, as shown in Table a,b and c.

\emph{\textbf{Table 3a}: Betweenness Centrality table}

\begin{longtable}[]{@{}
  >{\raggedright\arraybackslash}p{(\columnwidth - 4\tabcolsep) * \real{0.3333}}
  >{\raggedright\arraybackslash}p{(\columnwidth - 4\tabcolsep) * \real{0.3333}}
  >{\raggedright\arraybackslash}p{(\columnwidth - 4\tabcolsep) * \real{0.3333}}@{}}
\toprule
\endhead
ID & Degree & \begin{minipage}[t]{\linewidth}\raggedright
Betweenness\\
Centrality\strut
\end{minipage} \\
\bottomrule
MYC & 41 & 0.169411867 \\
TP53 & 39 & 0.146407871 \\
STAT3 & 40 & 0.13781554 \\
SNAI1 & 18 & 0.069823653 \\
NANOG & 25 & 0.066742667 \\
KLF4 & 15 & 0.055090028 \\
AR & 21 & 0.052755056 \\
POU5F1 & 21 & 0.047753251 \\
FOXM1 & 15 & 0.045555903 \\
FOXO3 & 16 & 0.040374857 \\
\bottomrule
\end{longtable}

\emph{\textbf{Table 3b}: Closeness Centrality table}

\begin{longtable}[]{@{}
  >{\raggedright\arraybackslash}p{(\columnwidth - 4\tabcolsep) * \real{0.3333}}
  >{\raggedright\arraybackslash}p{(\columnwidth - 4\tabcolsep) * \real{0.3333}}
  >{\raggedright\arraybackslash}p{(\columnwidth - 4\tabcolsep) * \real{0.3333}}@{}}
\toprule
\endhead
ID & Degree & \begin{minipage}[t]{\linewidth}\raggedright
Closeness\\
Centrality\strut
\end{minipage} \\
\bottomrule
MYC & 41 & 0.353760708 \\
STAT3 & 40 & 0.353404453 \\
TP53 & 39 & 0.348145458 \\
FOXO3 & 16 & 0.328586725 \\
NANOG & 25 & 0.328279347 \\
SNAI1 & 18 & 0.326751045 \\
POU5F1 & 21 & 0.315869147 \\
FOXM1 & 15 & 0.315301547 \\
FOXO1 & 15 & 0.312215856 \\
AR & 21 & 0.310833146 \\
\bottomrule
\end{longtable}

\emph{\textbf{Table 3c}: EigenVector Centrality table}

\begin{longtable}[]{@{}
  >{\raggedright\arraybackslash}p{(\columnwidth - 4\tabcolsep) * \real{0.3333}}
  >{\raggedright\arraybackslash}p{(\columnwidth - 4\tabcolsep) * \real{0.3333}}
  >{\raggedright\arraybackslash}p{(\columnwidth - 4\tabcolsep) * \real{0.3333}}@{}}
\toprule
\endhead
ID & Degree & \begin{minipage}[t]{\linewidth}\raggedright
EigenVector\\
Centrality\strut
\end{minipage} \\
\bottomrule
STAT3 & 40 & 0.366485503 \\
MYC & 41 & 0.303674676 \\
TP53 & 39 & 0.260577147 \\
NANOG & 25 & 0.240092313 \\
POU5F1 & 21 & 0.218315088 \\
FOXO3 & 16 & 0.173948202 \\
SOX2 & 16 & 0.173898883 \\
SNAI1 & 18 & 0.165160471 \\
KLF5 & 12 & 0.154141463 \\
FOXM1 & 15 & 0.153938826 \\
\bottomrule
\end{longtable}

MYC, TP53, and STAT3 were the top three TFs based on all three
centrality measurements. In terms of Betweenness centrality, MYC had the
highest score of 0.1694, followed by TP53 with a score of 0.1464 and
STAT3 with a score of 0.1378. Based on Closeness centrality, MYC had the
highest score of 0.3538, followed by STAT3 with a score of 0.3534 and
TP53 with a score of 0.3481. Based on eigenvector centrality, STAT3 had
the highest score of 0.3665, followed by MYC with a score of 0.3037 and
TP53 with a score of 0.2606.

The top three TFs identified in this study, MYC, TP53, and STAT3, are
well-known regulators of various cellular processes, including cell
growth, proliferation, and apoptosis. MYC is a proto-oncogene that plays
a critical role in regulating cell growth and proliferation. TP53 is a
tumor suppressor gene that is involved in regulating apoptosis and DNA
repair. STAT3 is a TF that plays a crucial role in regulating immune
responses and cell proliferation. The identification of these TFs as the
top regulators in the network supports their importance in regulating
cellular processes.

\hypertarget{pou5f1-summary}{%
\subparagraph{\texorpdfstring{Analyzing the prognostic relevance of POU5F1
}{Analyzing the prognostic relevance of POU5F1}}\label{pou5f1-summary}}
The network analysis of TF cascades dataset has identified POU5F1 as a key node. The TF POU5F1, known as OCT4, plays a crucial role in maintaining the essential characteristics of Embryonic Stem Cells(ESCs), such as their ability to renew themselves and their pluripotency, which is the capacity to develop into different cell types\cite{Stower2013SuperEnhancers, Wang2015TheCells}. Together with other key transcription factors, SOX2 and NANOG, POU5F1 activates specific genes that are vital for keeping the ESCs' unique properties\cite{Stower2013SuperEnhancers}. This group of factors works by binding to specific DNA regions that control gene activity, thus maintaining the cells' identity and potential for differentiation. Alterations in the function or levels of POU5F1 can disrupt these processes and may contribute to the development of cancer, as this gene can act as an oncogene when improperly regulated. Given its central role in cell identity and proliferation, POU5F1 is a subject of intense study, not only for its fundamental role in cell biology but also for its potential as a target in cancer treatment strategies.

\includegraphics{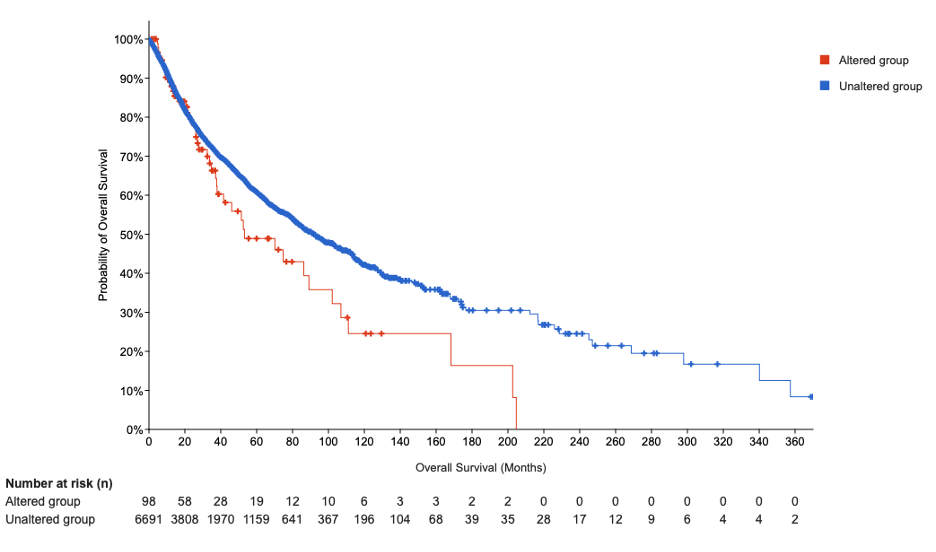}

\emph{\textbf{Figure 10} - KM Curve}

\emph{\textbf{Table 4a:} Survival Type Analysis}
\begin{longtable}{@{}p{\dimexpr(\textwidth-8\tabcolsep)/4\relax}p{\dimexpr(\textwidth-8\tabcolsep)/4\relax}p{\dimexpr(\textwidth-8\tabcolsep)/4\relax}p{\dimexpr(\textwidth-8\tabcolsep)/4\relax}@{}}
\hline
Survival Type & Number of Patients & p-Value & q-Value \\
\hline
\endfirsthead
\multicolumn{4}{c}%
{\tablename\ \thetable\ -- \textit{Continued from previous page}} \\
\hline
Survival Type & Number of Patients & p-Value & q-Value \\
\hline
\endhead
\hline
\multicolumn{4}{r}{\textit{Continued on next page}} \\
\endfoot
\hline
\endlastfoot
Disease-specific & 6553 & 4.188e-3 & 0.0168 \\
Overall & 6789 & 0.0318 & 0.0437 \\
Progression Free & 6787 & 0.0328 & 0.0437 \\
Disease Free & 3736 & 0.604 & 0.604 \\
\end{longtable}

\emph{\textbf{Table 4b:} Survival plot summary} 
\begin{longtable}{@{}p{\dimexpr(\textwidth-8\tabcolsep)/4\relax}p{\dimexpr(\textwidth-8\tabcolsep)/4\relax}p{\dimexpr(\textwidth-8\tabcolsep)/4\relax}p{\dimexpr(\textwidth-8\tabcolsep)/4\relax}@{}}
\hline
 & Number of Cases, Total & Number of Events & Median Months Overall \\
\hline
\endfirsthead
\multicolumn{4}{c}%
{\tablename\ \thetable\ -- \textit{Continued from previous page}} \\
\hline
 & Number of Cases, Total & Number of Events & Median Months Overall \\
\hline
\endhead
\hline
\multicolumn{4}{r}{\textit{Continued on next page}} \\
\endfoot
\hline
\endlastfoot
Altered group & 98 & 43 & 53.19 \\
Unaltered group & 6691 & 1946 & 91.99 \\
\end{longtable}

Subsequently, we performed Kaplan-Meier (KM) survival analyses(figure 9),which have provided valuable insights into the clinical significance of POU5F1 alterations in patient outcomes. The disease-specific survival analysis encompassed 6,553 patients, revealing a statistically significant disparity in survival outcomes based on POU5F1 status, with a p-value of 0.0041 and a q-value of 0.0168(Table 4a). This significance persisted even after adjusting for multiple survival comparisons, strengthening the evidence that POU5F1 alterations are indeed consequential. Furthermore, the overall survival analysis, including 6,789 patients, corroborated these findings with a p-value of 0.0318 and a q-value of 0.0437. Progression-free survival presented a similar pattern, suggesting a robust association between POU5F1 alterations and decreased survival metrics. In contrast, disease-free survival analysis of 3,736 patients did not exhibit a significant difference between altered and unaltered POU5F1 groups, indicating the specificity of POU5F1's impact on survival post-treatment.

The survival plot for the altered group, with 98 cases and 43 events, delineated a median overall survival of 53.19 months, substantively lower than the 91.99 months observed in the unaltered group of 6,691 patients(Table 4b). This substantial reduction in median survival time for patients with altered POU5F1 expression highlights the transcription factor's potential role as a prognostic marker.

The findings delineated above validate our initial hypothesis that POU5F1 alterations could portend unfavorable clinical outcomes. Given the pivotal role of POU5F1 in the regulatory networks of gene expression, its dysregulation may serve as a critical indicator of disease progression and patient prognosis. The statistical rigor of the KM survival analysis lends weight to the proposition that POU5F1 could be a viable target for therapeutic intervention, particularly in conditions where its expression is aberrant.

The confluence of network analysis and KM survival analysis in this study provides a compelling narrative about the role of POU5F1 in disease etiology and progression. It underscores the utility of this new dataset in unearthing potential therapeutic targets, with POU5F1 emerging as a TF of considerable interest for future research and clinical exploration.

\hypertarget{d-pathway-enrichment}{%
\subsubsection{3.2.2 d) Pathway Enrichment}\label{d-pathway-enrichment}}

Our pathway enrichment analysis provided valuable insights into the
biological functions and mechanisms of action of TFs. The enrichment
analysis resulted in functional categories that are significantly
enriched in the input gene set, which were ranked by the adjusted
p-value. The adjusted p-value is calculated using the Benjamini-Hochberg
method to control for the false discovery rate. In addition to the list
of enriched pathways, the enrichment analysis provided a variety of
other information and resources, such as gene-gene interactions, and
odds ratios.

\emph{\textbf{Table 5:} Result of pathway enrichment analysis}

\begin{longtable}[]{@{}
  >{\raggedright\arraybackslash}p{(\columnwidth - 12\tabcolsep) * \real{0.1890}}
  >{\raggedright\arraybackslash}p{(\columnwidth - 12\tabcolsep) * \real{0.0951}}
  >{\raggedright\arraybackslash}p{(\columnwidth - 12\tabcolsep) * \real{0.0950}}
  >{\raggedright\arraybackslash}p{(\columnwidth - 12\tabcolsep) * \real{0.1348}}
  >{\raggedright\arraybackslash}p{(\columnwidth - 12\tabcolsep) * \real{0.1225}}
  >{\raggedright\arraybackslash}p{(\columnwidth - 12\tabcolsep) * \real{0.1745}}
  >{\raggedright\arraybackslash}p{(\columnwidth - 12\tabcolsep) * \real{0.1892}}@{}}
\toprule
\endhead
Pathway & P-value & Z-score & Combined score & Adjusted p-value &
cascade\_value & cascade \\
\bottomrule
\begin{minipage}[t]{\linewidth}\raggedright
neuroactive ligand receptor\\
interaction\strut
\end{minipage} & 0.4375 & 1.7699 & 1.4633 & 0.4375 & 6727 & {[}'STAT3',
'JUNB', 'JUN', 'NANOG', 'POU5F1', .... {]} \\
focal adhesion & 0.371 & 2.2013 & 2.1829 & 0.3838 & 6928 & {[}'NCOA3',
'MYC', 'SNAI1', 'NANOG', 'POU5F1', '....{]} \\
mapk signaling pathway & 0.37 & 2.2173 & 2.2043 & 0.37 & 7022 &
{[}'EPAS1', 'SOX2', 'NANOG', 'POU5F1', 'SALL4'{]} \\
purine metabolism & 0.2848 & 3.0549 & 3.8371 & 0.3026 & 7284 & {[}'AR',
'MYC', 'SNAI1', 'NANOG', 'POU5F1', ....{]} \\
jak stat signaling pathway & 0.2897 & 2.9943 & 3.7097 & 0.2982 & 4750 &
{[}'LHX9', 'NR5A1', 'NR0B1', 'POU5F1', 'SALL4', .....{]} \\
axon guidance & 0.2619 & 3.3748 & 4.5212 & 0.2916 & 7530 & {[}'STAT3',
'SALL4', 'NANOG', 'POU5F1', 'SNAI1', ....{]} \\
\begin{minipage}[t]{\linewidth}\raggedright
natural killer cell mediated\\
cytotoxicity\strut
\end{minipage} & 0.2673 & 3.2952 & 4.3481 & 0.2916 & 7530 & {[}'STAT3',
'SALL4', 'NANOG', 'POU5F1', 'SNAI1',....{]} \\
tight junction & 0.266 & 3.3149 & 4.39 & 0.2821 & 5310 & {[}'OTX2',
'HMGA2', 'TWIST1', 'POU5F1', 'SALL4',....{]} \\
cell cycle & 0.2216 & 4.1002 & 6.178 & 0.2574 & 7199 & {[}'STAT3',
'NFATC2', 'SOX2', 'NANOG', 'POU5F1',....{]} \\
melanogenesis & 0.2141 & 4.2667 & 6.577 & 0.2497 & 6966 & {[}'STAT6',
'PPARG', 'TP53', 'NANOG', 'POU5F1', ....{]} \\
gnrh signaling pathway & 0.2064 & 4.4473 & 7.0166 & 0.2492 & 6779 &
{[}'NCOA1', 'MYC', 'SNAI1', 'NANOG', 'POU5F1', ....{]} \\
pyrimidine metabolism & 0.1831 & 5.0933 & 8.6465 & 0.2273 & 7199 &
{[}'STAT3', 'NFATC2', 'SOX2', 'NANOG', 'POU5F1', ....{]} \\
apoptosis & 0.1772 & 5.2851 & 9.1462 & 0.2215 & 7480 & {[}'EGR2',
'CEBPB', 'PPARG', 'TP53', 'NANOG', ....{]} \\
small cell lung cancer & 0.1851 & 5.0324 & 8.4893 & 0.217 & 6528 &
{[}'SNAI2', 'RUNX2', 'SP7', 'SATB2', 'NANOG', ....{]} \\
erbb signaling pathway & 0.1851 & 5.0324 & 8.4893 & 0.217 & 6528 &
{[}'SNAI2', 'RUNX2', 'SP7', 'SATB2', 'NANOG', ....{]} \\
chronic myeloid leukemia & 0.1652 & 5.7154 & 10.2914 & 0.216 & 6956 &
{[}'HOXA5', 'SOX2', 'NANOG', 'POU5F1', 'SALL4', ....{]} \\
glioma & 0.1386 & 6.9379 & 13.71 & 0.2141 & 7314 & {[}'STAT3', 'FOS',
'SNAI1', 'NANOG', 'POU5F1', ....{]} \\
adherens junction & 0.1652 & 5.7154 & 10.2914 & 0.2141 & 7314 &
{[}'STAT3', 'FOS', 'SNAI1', 'NANOG', 'POU5F1', ....{]} \\
basal cell carcinoma & 0.124 & 7.84 & 16.3687 & 0.2141 & 7314 &
{[}'STAT3', 'FOS', 'SNAI1', 'NANOG', 'POU5F1', ....{]} \\
melanoma & 0.1571 & 6.0432 & 11.1852 & 0.2141 & 7314 & {[}'STAT3',
'FOS', 'SNAI1', 'NANOG', 'POU5F1', ....{]} \\
\bottomrule
\end{longtable}

The results of pathway enrichment analysis using Enrichr revealed
significant enrichment of various pathways and functional categories
among the TF gene set. For the 81,488 cascades, we obtained 2 million
pathways with an average of 25 pathways for each cascade(Table 5). The
top enriched pathways included "Transcriptional misregulation in
cancer," "gnrh signaling pathway," and "Th17 cell differentiation,"
which are known to be involved in cancer and various other diseases.
Other enriched pathways included " neuroactive ligand receptor
interaction," "adipocytokine signaling pathway," and "hedgehog signaling
pathway," which are involved in the development, differentiation, and
cell signaling.

\hypertarget{knowledge-graph}{%
\subsubsection{3.2.3 Knowledge Graph}\label{knowledge-graph}}

The knowledge graph is constructed as an undirected multigraph with a
total of 9,784 nodes and 38,627 edges(Table 6). The node types present in the graph
include only a single type, i.e., source, which represents the TFs in
the TF cascade. Each TF node points towards the next TF node in the
cascade and also towards the pathways, which are the targets in this
graph. The graph is node-labeled and attributed, where each node
contains information about the P-value, Z-score, and combined score for
all the available pathways.

Furthermore, the edges in the graph are unlabelled but attributed, and
the edge type present in the graph is source-target-\textgreater source.
The weights of the edges range from 1.13e-19 to 0.437467, with a mean of
0.154782 and standard deviation of 0.114678. The graph can be used to
represent the relationships between the TFs and pathways in the cascade
and can be leveraged for further analysis and prediction tasks.

\emph{\textbf{Table 6:} Graph Summary}

\begin{longtable}[]{@{}
  >{\raggedright\arraybackslash}p{(\columnwidth - 4\tabcolsep) * \real{0.3333}}
  >{\raggedright\arraybackslash}p{(\columnwidth - 4\tabcolsep) * \real{0.3334}}
  >{\raggedright\arraybackslash}p{(\columnwidth - 4\tabcolsep) * \real{0.3334}}@{}}
\toprule
\endhead
\textbf{Node Type} & \textbf{Description} & \textbf{Frequency} \\
\bottomrule
TFs & The final TF in a series of TF cascades is connected in a
unidirectional manner to the preceding TFs within the cascade. & 426 \\
Pathways & Each terminal TF cascade will be linked to its corresponding
pathway. & 49 \\
Total nodes & & 475 \\
\bottomrule
\end{longtable}

\begin{longtable}[]{@{}
  >{\raggedright\arraybackslash}p{(\columnwidth - 4\tabcolsep) * \real{0.3333}}
  >{\raggedright\arraybackslash}p{(\columnwidth - 4\tabcolsep) * \real{0.3334}}
  >{\raggedright\arraybackslash}p{(\columnwidth - 4\tabcolsep) * \real{0.3334}}@{}}
\toprule
\endhead
\textbf{Edges} & \textbf{Description} & \textbf{Frequency} \\
\bottomrule
All of TFs & Weights for connections between transcriptions in each
cascade are assigned as 1, and for the entire graph, it would be the sum
of all connections in the dataset. & 5209 \\
All pairs of TFs and Pathways & Weights Information on P-value, Z-score,
combined score for all available Pathways. The pairs without connection
are marked 0 & 33418 \\
Total edges & & 38627 \\
\bottomrule
\end{longtable}

\textbf{3.2.4 GraphML on the Enriched TF Cascades Knowledge Graph}

The prediction of the next TF in a cascade was modeled as a supervised
learning problem on top of TF node representations. We first generated a
sub-graph where each node represents a transcription factor and the
edges represent the relationship between the TFs. To perform
similarity-based TF link prediction, we employed a network embedding
technique called Node2Vec, which calculates node embeddings using random
walk. The Node2Vec algorithm first runs random walks on the graph to
obtain context pairs, which are then used to train a Word2Vec model. The
resulting embeddings are learned in such a way to ensure that nodes that
are close in the graph remain close in the embedding space. The
parameters used for Node2Vec in our study were number of walks = 10,
length of each random walk = 80, number of iterations = 10. we obtained
97,840 random walks.

Once the node embeddings are obtained, we can use them to perform link
prediction using various binary operators such as Hadamard, L1, L2, and
average. These operators are applied on the embeddings of the source and
target nodes of each sampled edge to calculate link embeddings for the
positive and negative edge samples. We used a random 75:25 split, with
75\% of the data used for training the model and the remaining 25\% used
for testing. This ensures that the model is trained on a sufficiently
large dataset while still having enough data left for testing. We then
use a logistic regression classifier to train on the embeddings of the
positive and negative examples to predict whether an edge between two
nodes should exist or not. We evaluate the performance of the link
classifier for each of the four operators on the training data with node
embeddings calculated on the graph and select the best classifier.

\emph{\textbf{Table 7:} TF Graph ML ROC-AUC scores}

\begin{longtable}[]{@{}
  >{\raggedright\arraybackslash}p{(\columnwidth - 2\tabcolsep) * \real{0.5221}}
  >{\raggedright\arraybackslash}p{(\columnwidth - 2\tabcolsep) * \real{0.4779}}@{}}
\toprule
\endhead
\textbf{Operator} & \textbf{ROC-AUC score} \\
\bottomrule
Hadamard & 0.800292 \\
L1 & 0.798369 \\
L2 & 0.803425 \\
Average & 0.971989 \\
\bottomrule
\end{longtable}

We considered four different operators, namely Hadamard, L1, L2, and
Average operators to generate link embeddings before
classification(Table 7). Our experimental results showed that the best
operator was the Average operator with an ROC AUC score of 0.971989. The choice of operator has implications for the quality of TF cascade link prediction and can be explored further to optimize the performance of the link prediction model.

\hypertarget{discussion}{%
\subsection{\texorpdfstring{4. Discussion
}{4. Discussion }}\label{discussion}}

Transcription factors are crucial regulators of gene expression and play
a vital role in various cellular processes. In this study, we
constructed a compendium of TF cascades using data extracted from the
STRING database. Our analysis revealed 81,488 unique TF cascades, with
the longest cascade consisting of 62 TFs, highlighting the intricate
nature of TF interactions and the collaborative efforts of multiple TFs
in gene regulation.

By employing centrality measurements, we identified the top 10 TFs with
the highest regulatory influence. These TFs represent key players in the
network and provide valuable insights for researchers interested in
studying specific TFs. Understanding the regulatory roles and impact of
these highly influential TFs is crucial for gaining deeper insights into
their functional significance and their potential implications in
disease processes.

One significant application of the PageRank and graphML results obtained
from our analysis is the identification of potential drug targets among
the highly ranked proteins. Discovering new drug targets is essential
for developing effective treatments for cancer and other diseases.
Leveraging the PageRank results, researchers can explore previously
uninvestigated proteins as potential drug targets. This approach serves
as a valuable checklist to guide the search for novel therapeutic
interventions.

Furthermore, our pathway enrichment analysis uncovered significant
enrichment of various pathways and functional categories, including
those involved in cancer, development, differentiation, and cell
signaling. These findings provide valuable insights into the
dysregulation of TFs in disease states. The enriched pathways identified
in this study may serve as potential targets for therapeutic
intervention, offering new avenues for drug discovery and treatment
strategies for diseases associated with TF dysregulation.

To ensure the accessibility and usability of our findings, we have made
the TF cascades dataset, knowledge graph, and graphML methods publicly
available. Additionally, we have developed a dedicated website for
researchers to access and utilize this resource for their
investigations. This resource serves as a valuable tool for researchers
seeking to comprehend the complex network of TF interactions and their
regulatory roles in cellular processes.

\emph{\textbf{Clinical Perspective and Validation: How identification of TF Cascades can allow identify therapeutic targets}}

Oncogenesis, post-operative wound healing, and hormonal aberrancies leading to metabolic derangements are often a result of improper TF cascade signaling\cite{Kciuk2022Cancer-associatedResponse, Zhang2020CurrentFormation,Ku2020MasterCancer}. Determinants of TF to bind to DNA to concord to a particular function is based on the structure of motifs, of which 80\% are helix–turn–helix (HTH), zinc finger (ZF), leucine zipper, forkhead, or helix–loop–helix (HLH). Motif structures were earlier used to classify TFs, and each class proved a variable set of functions. Notably, HTH motifs were associated with hematopoietic precursor cells and leukemias\cite{Shima2011DeregulatedLeukemia}, whereas, Krüppel-like factors (KLFs) belong to multiple biological processes and are involved also in carcinogenesis\cite{McConnell2010MammalianDiseases}. On the other hand, FOXO genes - that belong to the fork-head box family – produce tumor suppressor proteins that are often inactivated during human cancers.

Of the enriched pathways identified in our study, NANOG was seen to be of highest PageRank, followed by TP53. NANOG, a homeobox transcription factor essential for maintaining the pluripotency of embryonic stem cells (ESCs), plays a critical role in tumorigenesis, progression, and metastasis in a variety of human cancers. The self-renewal and survival of cancer stem cells (CSCs), responsible for tumor initiation, is promoted by NANOG\cite{Najafzadeh2021TheMediator}. Creating targeted treatments for NANOG has drawn more attention in recent years. Understanding NANOG's role may lead to it becoming a therapeutic target to halt cancer progression\cite{Vasefifar2022NanogProgression}. Recent studies have shown NANOG has demonstrated synergistic induction of naïve pluripotent cells through LIF signal transduction, causing increased STAT3 whereas NANOG limits the effect of KLF4\cite{Stuart2014NANOGProgram}. 

Previous studies have demonstrated that Krupple-like factor 5 (KLF5), a zinc-finger transcriptional factor, may be a potential therapeutic target for oxaliplatin-resistant CRC (colorectal carcinoma). KLF also activates gene promoters PDGF‐A/‐B, iNOS, PAI‐1, and VEGF receptors, making it a target for cardiovascular remodeling\cite{Nagai2005SignificanceRemodeling}. In squamous cell carcinoma, KLF5 positively regulates Sox4 expression and KLF5/Sox4 regulatory signaling facilitates tumorigenesis\cite{Li2014Kruppel-likeSox4}. Furthermore, Inhibition of KLF5 using siRNA or ML264 markedly decreased invasion and migration in EOC (Epithelial ovarian cancer) cells\cite{Siraj2020Krupple-LikeCancer}. However, Genes regulated by KLF5 have not been well characterized with the advent of TF cascading, conventional pharmacotherapies may be combined with KLF5 inhibition in the future. Sex-determining region Y box 2 (SOX2)TF— significant in embryonic development— is a key regulator of CSC’s in glioblastoma (GBM), malignant brain tumor with dismal prognosis. The current therapeutic target available for GBM is temozolomide (TMZ)\cite{Garros-Regulez2016TargetingGlioblastoma}. Increased SOX2 expression increases the resilience to SOX2, whereas its inhibition improves drug action. SOX2 inhibition by miRNA145 decreases the chemoresistance and improves TMZ sensitivity\cite{Siraj2020Krupple-LikeCancer}. 

MYC upregulates POU5F1(OCT4) attributing to cell renewal process. MYC interacts with NANOG and SOX2 to maintain the balance between stemness and differentiation of cells. This is imperative as the balance of the TF levels was not earlier accounted for the same property, rather it was considered to be solely by POU5F1\cite{Das2019MYCCells}. There is existence of a stoichiometric effect for SOX2, POU5F1, NANOG, MYC and KLF4, in regulating POU5F1 transcription\cite{Wang2012ThePOU5F1}. Our paper emphasizes the earlier known literature on TF functions and progressive knowledge of TF interactions impacting deeper understanding. Our program would allow to uncover further TF cascades that could later reveal pertinent interactions.

\emph{\textbf{Future Work}}

An intriguing finding in our dataset is the role of POU5F1, which is relatively understudied, despite being involved in 1,546 cascades. Further investigation can be conducted to assess the impact of specific
TFs, such as POU5F1 and NANOG, in cancers occurring in the testis and lung.  Analyzing their involvement and regulatory influence in these specific
anatomical regions can provide insights into their potential
contributions to cancer development and progression.In addition, the
application of a smarter PageRank-like algorithm, incorporating
disease-specific information and weighting incoming edges based on
protein importance, can enhance the analysis and prioritize potential
drug targets more accurately.

Furthermore, probabilistic models can be employed to approximate the
likelihood of different outcomes following mutations or alterations in
TF cascades. By incorporating specific data for each protein within the
cascade, researchers can gain a deeper understanding of the potential
consequences, such as cancer development, activation of other TFs, or
gene dysregulation\cite{Shameer2017TranslationalStreams}.

Forward, TF research will enable precision medicine, by obtaining a deeper understanding of transcription factor networks, creating personalized treatment approaches\cite{Tan2019Network-basedParadigm}. TF networks would also allow drug repurposing\cite{Shameer2018SystematicRepositioning}, tissue engineering and unfolding regenerative medicine by developing specific cell types from stem cells\cite{Grath2019DirectMedicine}.  Transcription factors can be attractive drug targets\cite{Butt1995TranscriptionSelectivity}. With emerging CRISPR-Cas9 use for genome editing, the understanding of TF sites would enable cellular reprogramming\cite{Pandelakis2020CRISPR-BasedProgramming}. Involving complex machine learning models on TFs’ would allow to predict its impact on health and disease. Moreover, as of precise gene manipulations of gene expression through TFs, social, legal and ethical considerations must be taken into account in the future with careful oversight and guidelines\cite{Moradi2019ResearchConsiderations}.

\hypertarget{conclusion}{%
\subsection{5. Conclusion}\label{conclusion}}

This study contributes to the understanding of the complex network of TF
interactions and their regulatory roles in cellular processes. By
constructing a comprehensive compendium of TF cascades and identifying
highly influential TFs, we provide valuable insights for researchers
interested in studying specific TFs. The identification of potential
drug targets among the highly ranked proteins offers new opportunities
for therapeutic interventions in cancer and other diseases.
Additionally, our pathway enrichment analysis sheds light on
dysregulated pathways associated with TFs, highlighting potential
targets for therapeutic intervention.

We have made the TF cascades dataset, knowledge graph, and graphML
methods publicly available, along with a dedicated website, ensuring the
accessibility and usability of this resource for the research community.
This compendium serves as a valuable tool for investigating the
intricate interplay of TFs and their roles in cellular processes, aiding
researchers in the quest for novel therapeutic targets and treatments
for diseases associated with TF dysregulation.


\bibliographystyle{vancouver}
\bibliography{references}

\begin{thebibliography}{10}

\bibitem{Lambert2018TheFactors}
Lambert SA, Jolma A, Campitelli LF, Das PK, Yin Y, Albu M, et~al.
\newblock {The human transcription factors}.
\newblock Cell. 2018;172(4):650-65.

\bibitem{Lee2020GlobalStudy}
Lee HA, Kung HH, Lee YJ, Chao JCJ, Udayasankaran JG, Fan HC, et~al.
\newblock {Global infectious disease surveillance and case tracking system for
  COVID-19: development study}.
\newblock JMIR Medical Informatics. 2020;8(12):e20567.

\bibitem{Lee2013TranscriptionalDisease}
Lee TI, Young RA.
\newblock {Transcriptional regulation and its misregulation in disease}.
\newblock Cell. 2013;152(6):1237-51.

\bibitem{Fong2013SkeletalRe-programming}
Fong AP, Tapscott SJ.
\newblock {Skeletal muscle programming and re-programming}.
\newblock Current opinion in genetics {\&} development. 2013;23(5):568-73.

\bibitem{Takahashi2016APluripotency}
Takahashi K, Yamanaka S.
\newblock {A decade of transcription factor-mediated reprogramming to
  pluripotency}.
\newblock Nature reviews Molecular cell biology. 2016;17(3):183-93.

\bibitem{Naika2013STIFDB2:Rice}
Naika M, Shameer K, Mathew OK, Gowda R, Sowdhamini R.
\newblock {STIFDB2: an updated version of plant stress-responsive transcription
  factor database with additional stress signals, stress-responsive
  transcription factor binding sites and stress-responsive genes in Arabidopsis
  and rice}.
\newblock Plant Cell Physiol. 2013;54(2):e8.
\newblock Available from: \url{https://www.ncbi.nlm.nih.gov/pubmed/23314754}.

\bibitem{Phillips2008TheExpression}
Phillips T.
\newblock {The role of methylation in gene expression}.
\newblock Nature Education. 2008;1(1):116.

\bibitem{Alvarez2020TransientCascade}
Alvarez JM, Schinke AL, Brooks MD, Pasquino A, Leonelli L, Varala K, et~al.
\newblock {Transient genome-wide interactions of the master transcription
  factor NLP7 initiate a rapid nitrogen-response cascade}.
\newblock Nature communications. 2020;11(1):1157.

\bibitem{Surget2013UncoveringPerspective}
Surget S, Khoury MP, Bourdon JC.
\newblock {Uncovering the role of p53 splice variants in human malignancy: a
  clinical perspective}.
\newblock OncoTargets and therapy. 2013:57-68.

\bibitem{Cerami2012TheData}
Cerami E, Gao J, Dogrusoz U, Gross BE, Sumer SO, Aksoy BA, et~al.
\newblock {The cBio cancer genomics portal: an open platform for exploring
  multidimensional cancer genomics data}.
\newblock Cancer discovery. 2012;2(5):401-4.

\bibitem{Karamouzis2011TranscriptionParadigm}
Karamouzis MV, Papavassiliou AG.
\newblock {Transcription factor networks as targets for therapeutic
  intervention of cancer: the breast cancer paradigm}.
\newblock Molecular Medicine. 2011;17:1133-6.

\bibitem{Zhang2017Network-basedOncology}
Zhang W, Chien J, Yong J, Kuang R.
\newblock {Network-based machine learning and graph theory algorithms for
  precision oncology}.
\newblock NPJ precision oncology. 2017;1(1):25.

\bibitem{Olayan2018DDR:Approaches}
Olayan RS, Ashoor H, Bajic VB.
\newblock {DDR: efficient computational method to predict drug–target
  interactions using graph mining and machine learning approaches}.
\newblock Bioinformatics. 2018;34(7):1164-73.

\bibitem{Thafar2022Affinity2Vec:Learning}
Thafar MA, Alshahrani M, Albaradei S, Gojobori T, Essack M, Gao X.
\newblock {Affinity2Vec: drug-target binding affinity prediction through
  representation learning, graph mining, and machine learning}.
\newblock Scientific reports. 2022;12(1):4751.

\bibitem{Szklarczyk2016TheAccessible}
Szklarczyk D, Morris JH, Cook H, Kuhn M, Wyder S, Simonovic M, et~al.
\newblock {The STRING database in 2017: quality-controlled protein–protein
  association networks, made broadly accessible}.
\newblock Nucleic acids research. 2016:gkw937.

\bibitem{Palasca2018TISSUESExpression}
Palasca O, Santos A, Stolte C, Gorodkin J, Jensen LJ.
\newblock {TISSUES 2.0: an integrative web resource on mammalian tissue
  expression}.
\newblock Database. 2018;2018:bay003.

\bibitem{Consortium2019UniProt:Knowledge}
Consortium U.
\newblock {UniProt: a worldwide hub of protein knowledge}.
\newblock Nucleic acids research. 2019;47(D1):D506-15.

\bibitem{Jacobsen2015Cgdsr:CGDS}
Jacobsen A.
\newblock {cgdsr: R-based API for accessing the MSKCC cancer genomics data
  server (CGDS)}.
\newblock R package version. 2015;1(5).

\bibitem{Page1998TheReport}
Page L.
\newblock {The pagerank citation ranking: Bringing order to the web. Technical
  report}.
\newblock Stanford Digital Library Technologies Project, 1998. 1998.

\bibitem{Shameer2018ADisease}
Shameer K, Dow G, Glicksberg BS, Johnson KW, Ze Y, Tomlinson MS, et~al.
\newblock {A Network-Biology Informed Computational Drug Repositioning Strategy
  to Target Disease Risk Trajectories and Comorbidities of Peripheral Artery
  Disease}.
\newblock AMIA Jt Summits Transl Sci Proc. 2018;2017:108-17.
\newblock Available from: \url{https://www.ncbi.nlm.nih.gov/pubmed/29888052}.

\bibitem{Glicksberg2016ComparativeNetworks}
Glicksberg BS, Li L, Badgeley MA, Shameer K, Kosoy R, Beckmann ND, et~al.
\newblock {Comparative analyses of population-scale phenomic data in electronic
  medical records reveal race-specific disease networks}.
\newblock Bioinformatics. 2016;32(12):i101-10.
\newblock Available from: \url{https://www.ncbi.nlm.nih.gov/pubmed/27307606}.

\bibitem{Kuleshov2016Enrichr:Update}
Kuleshov MV, Jones MR, Rouillard AD, Fernandez NF, Duan Q, Wang Z, et~al.
\newblock {Enrichr: a comprehensive gene set enrichment analysis web server
  2016 update}.
\newblock Nucleic acids research. 2016;44(W1):W90-7.

\bibitem{Thissen2002QuickComparisons}
Thissen D, Steinberg L, Kuang D.
\newblock {Quick and easy implementation of the Benjamini-Hochberg procedure
  for controlling the false positive rate in multiple comparisons}.
\newblock Journal of educational and behavioral statistics. 2002;27(1):77-83.

\bibitem{Li2017ComprehensiveDatasets}
Li WX, He K, Tang L, Dai SX, Li GH, Lv WW, et~al.
\newblock {Comprehensive tissue-specific gene set enrichment analysis and
  transcription factor analysis of breast cancer by integrating 14 gene
  expression datasets}.
\newblock Oncotarget. 2017;8(4):6775.

\bibitem{Yacoumatos2021TrialGraph:Trials}
Yacoumatos C, Bragaglia S, Kanakia A, Svang{\aa}rd N, Mangion J, Donoghue C,
  et~al.
\newblock {TrialGraph: Machine Intelligence Enabled Insight from Graph
  Modelling of Clinical Trials}.
\newblock arXiv preprint arXiv:211208211. 2021.

\bibitem{Hagberg2008ExploringNetworkX}
Hagberg A, Swart P, S~Chult D.
\newblock {Exploring network structure, dynamics, and function using NetworkX};
  2008.

\bibitem{Gaudelet2021UtilizingDevelopment}
Gaudelet T, Day B, Jamasb AR, Soman J, Regep C, Liu G, et~al.
\newblock {Utilizing graph machine learning within drug discovery and
  development}.
\newblock Briefings in bioinformatics. 2021;22(6):bbab159.

\bibitem{Makarov2021SurveyGraphs}
Makarov I, Kiselev D, Nikitinsky N, Subelj L.
\newblock {Survey on graph embeddings and their applications to machine
  learning problems on graphs}.
\newblock PeerJ Computer Science. 2021;7:e357.

\bibitem{Stower2013SuperEnhancers}
Stower H.
\newblock {Super enhancers}.
\newblock Nature reviews Genetics. 2013;14(6):367.

\bibitem{Wang2015TheCells}
Wang YJ, Herlyn M.
\newblock {The emerging roles of Oct4 in tumor-initiating cells}.
\newblock American Journal of Physiology-Cell Physiology. 2015;309(11):C709-18.

\bibitem{Kciuk2022Cancer-associatedResponse}
Kciuk M, Gielecinska A, Kolat D, Kaluzinska Z, Kontek R.
\newblock {Cancer-associated transcription factors in DNA damage response}.
\newblock Biochim Biophys Acta Rev Cancer. 2022;1877(4):188757.
\newblock Available from: \url{https://www.ncbi.nlm.nih.gov/pubmed/35781034}.

\bibitem{Zhang2020CurrentFormation}
Zhang T, Wang XF, Wang ZC, Lou D, Fang QQ, Hu YY, et~al.
\newblock {Current potential therapeutic strategies targeting the TGF-beta/Smad
  signaling pathway to attenuate keloid and hypertrophic scar formation}.
\newblock Biomed Pharmacother. 2020;129:110287.
\newblock Available from: \url{https://www.ncbi.nlm.nih.gov/pubmed/32540643}.

\bibitem{Ku2020MasterCancer}
Ku HC, Cheng CF.
\newblock {Master Regulator Activating Transcription Factor 3 (ATF3) in
  Metabolic Homeostasis and Cancer}.
\newblock Front Endocrinol (Lausanne). 2020;11:556.
\newblock Available from: \url{https://www.ncbi.nlm.nih.gov/pubmed/32922364}.

\bibitem{Shima2011DeregulatedLeukemia}
Shima Y, Kitabayashi I.
\newblock {Deregulated transcription factors in leukemia}.
\newblock Int J Hematol. 2011;94(2):134-41.
\newblock Available from: \url{https://www.ncbi.nlm.nih.gov/pubmed/21823042}.

\bibitem{McConnell2010MammalianDiseases}
McConnell BB, Yang VW.
\newblock {Mammalian Kruppel-like factors in health and diseases}.
\newblock Physiol Rev. 2010;90(4):1337-81.
\newblock Available from: \url{https://www.ncbi.nlm.nih.gov/pubmed/20959618}.

\bibitem{Najafzadeh2021TheMediator}
Najafzadeh B, Asadzadeh Z, Motafakker~Azad R, Mokhtarzadeh A, Baghbanzadeh A,
  Alemohammad H, et~al.
\newblock {The oncogenic potential of NANOG: An important cancer induction
  mediator}.
\newblock J Cell Physiol. 2021;236(4):2443-58.
\newblock Available from: \url{https://www.ncbi.nlm.nih.gov/pubmed/32960465}.

\bibitem{Vasefifar2022NanogProgression}
Vasefifar P, Motafakkerazad R, Maleki LA, Najafi S, Ghrobaninezhad F,
  Najafzadeh B, et~al.
\newblock {Nanog, as a key cancer stem cell marker in tumor progression}.
\newblock Gene. 2022;827:146448.
\newblock Available from: \url{https://www.ncbi.nlm.nih.gov/pubmed/35337852}.

\bibitem{Stuart2014NANOGProgram}
Stuart HT, van Oosten AL, Radzisheuskaya A, Martello G, Miller A, Dietmann S,
  et~al.
\newblock {NANOG amplifies STAT3 activation and they synergistically induce the
  naive pluripotent program}.
\newblock Curr Biol. 2014;24(3):340-6.
\newblock Available from: \url{https://www.ncbi.nlm.nih.gov/pubmed/24462001}.

\bibitem{Nagai2005SignificanceRemodeling}
Nagai R, Suzuki T, Aizawa K, Shindo T, Manabe I.
\newblock {Significance of the transcription factor KLF5 in cardiovascular
  remodeling}.
\newblock J Thromb Haemost. 2005;3(8):1569-76.
\newblock Available from: \url{https://www.ncbi.nlm.nih.gov/pubmed/16102021}.

\bibitem{Li2014Kruppel-likeSox4}
Li Q, Dong Z, Zhou F, Cai X, Gao Y, Wang LW.
\newblock {Kruppel-like factor 5 promotes lung tumorigenesis through
  upregulation of Sox4}.
\newblock Cell Physiol Biochem. 2014;33(1):1-10.
\newblock Available from: \url{https://www.ncbi.nlm.nih.gov/pubmed/24401325}.

\bibitem{Siraj2020Krupple-LikeCancer}
Siraj AK, Pratheeshkumar P, Divya SP, Parvathareddy SK, Alobaisi KA, Thangavel
  S, et~al.
\newblock {Krupple-Like Factor 5 is a Potential Therapeutic Target and
  Prognostic Marker in Epithelial Ovarian Cancer}.
\newblock Front Pharmacol. 2020;11:598880.
\newblock Available from: \url{https://www.ncbi.nlm.nih.gov/pubmed/33424607}.

\bibitem{Garros-Regulez2016TargetingGlioblastoma}
Garros-Regulez L, Garcia I, Carrasco-Garcia E, Lantero A, Aldaz P,
  Moreno-Cugnon L, et~al.
\newblock {Targeting SOX2 as a Therapeutic Strategy in Glioblastoma}.
\newblock Front Oncol. 2016;6:222.
\newblock Available from: \url{https://www.ncbi.nlm.nih.gov/pubmed/27822457}.

\bibitem{Das2019MYCCells}
Das B, Pal B, Bhuyan R, Li H, Sarma A, Gayan S, et~al.
\newblock {MYC Regulates the HIF2alpha Stemness Pathway via Nanog and Sox2 to
  Maintain Self-Renewal in Cancer Stem Cells versus Non-Stem Cancer Cells}.
\newblock Cancer Res. 2019;79(16):4015-25.
\newblock Available from: \url{https://www.ncbi.nlm.nih.gov/pubmed/31266772}.

\bibitem{Wang2012ThePOU5F1}
Wang WP, Tzeng TY, Wang JY, Lee DC, Lin YH, Wu PC, et~al.
\newblock {The EP300, KDM5A, KDM6A and KDM6B chromatin regulators cooperate
  with KLF4 in the transcriptional activation of POU5F1}.
\newblock PLoS One. 2012;7(12):e52556.
\newblock Available from: \url{https://www.ncbi.nlm.nih.gov/pubmed/23272250}.

\bibitem{Shameer2017TranslationalStreams}
Shameer K, Badgeley MA, Miotto R, Glicksberg BS, Morgan JW, Dudley JT.
\newblock {Translational bioinformatics in the era of real-time biomedical,
  health care and wellness data streams}.
\newblock Brief Bioinform. 2017;18(1):105-24.
\newblock Available from: \url{https://www.ncbi.nlm.nih.gov/pubmed/26876889}.

\bibitem{Tan2019Network-basedParadigm}
Tan A, Huang H, Zhang P, Li S.
\newblock {Network-based cancer precision medicine: A new emerging paradigm}.
\newblock Cancer Lett. 2019;458:39-45.
\newblock Available from: \url{https://www.ncbi.nlm.nih.gov/pubmed/31125640}.

\bibitem{Shameer2018SystematicRepositioning}
Shameer K, Glicksberg BS, Hodos R, Johnson KW, Badgeley MA, Readhead B, et~al.
\newblock {Systematic analyses of drugs and disease indications in RepurposeDB
  reveal pharmacological, biological and epidemiological factors influencing
  drug repositioning}.
\newblock Brief Bioinform. 2018;19(4):656-78.
\newblock Available from: \url{https://www.ncbi.nlm.nih.gov/pubmed/28200013}.

\bibitem{Grath2019DirectMedicine}
Grath A, Dai G.
\newblock {Direct cell reprogramming for tissue engineering and regenerative
  medicine}.
\newblock J Biol Eng. 2019;13:14.
\newblock Available from: \url{https://www.ncbi.nlm.nih.gov/pubmed/30805026}.

\bibitem{Butt1995TranscriptionSelectivity}
Butt TR, Karathanasis SK.
\newblock {Transcription factors as drug targets: opportunities for therapeutic
  selectivity}.
\newblock Gene Expr. 1995;4(6):319-36.
\newblock Available from: \url{https://www.ncbi.nlm.nih.gov/pubmed/7549464}.

\bibitem{Pandelakis2020CRISPR-BasedProgramming}
Pandelakis M, Delgado E, Ebrahimkhani MR.
\newblock {CRISPR-Based Synthetic Transcription Factors In Vivo: The Future of
  Therapeutic Cellular Programming}.
\newblock Cell Syst. 2020;10(1):1-14.
\newblock Available from: \url{https://www.ncbi.nlm.nih.gov/pubmed/31972154}.

\bibitem{Moradi2019ResearchConsiderations}
Moradi S, Mahdizadeh H, Saric T, Kim J, Harati J, Shahsavarani H, et~al.
\newblock {Research and therapy with induced pluripotent stem cells (iPSCs):
  social, legal, and ethical considerations}.
\newblock Stem Cell Res Ther. 2019;10(1):341.
\newblock Available from: \url{https://www.ncbi.nlm.nih.gov/pubmed/31753034}.

\end{thebibliography}
\end{document}